\documentclass[prb,twocolumn]{revtex4-1}

\usepackage{amsmath}  
\usepackage{amsfonts} 
\usepackage{amssymb} 
\usepackage{graphicx} 
\usepackage{rotating} 
\usepackage{array}
\usepackage{comment}
\usepackage{color}
\usepackage[colorlinks,citecolor=blue]{hyperref}
\newcommand{\ngr}[1]{\boldsymbol{#1}}
\newcommand{\tun}{\mathbf{\hat{t}}}
\newcommand{\nun}{\mathbf{\hat{n}}}
\newcommand{\bun}{\mathbf{\hat{b}}}
\newcommand{\xun}{\boldsymbol{\hat{x}}}
\newcommand{\yun}{\boldsymbol{\hat{y}}}
\newcommand{\zun}{\boldsymbol{\hat{z}}}
\def\e{{\ e}}
\def\sech{\rm sech}
\def\sech{{\rm\ sech}}

\def\v{\mathbf}

\renewcommand{\phi}{\varphi}

\begin{document}


\title{Slipping on an Arbitrary Surface with Friction}

\author{Felipe Gonz\'alez-Cataldo}
 \email{fgonzalez@lpmd.cl}
\affiliation{Departamento de F\'isica, Facultad de Ciencias, Universidad de Chile, Casilla 653, Santiago, Chile.}
\author{Gonzalo Guti\'errez}
 \email{gonzalo@fisica.ciencias.uchile.cl}
\affiliation{Departamento de F\'isica, Facultad de Ciencias, Universidad de Chile, Casilla 653, Santiago, Chile.}
\author{Julio M. Y\'a\~{n}ez}%
 \email{juyanez@ucn.cl}
\affiliation{Departamento de F\'isica, Facultad de Ciencias, Universidad Cat\'olica del Norte}


\date{\today}

\begin{abstract}
The motion of a block slipping on a surface is a well studied problem for
flat and circular surfaces, but the necessary conditions for the block
to leave (or not) the surface deserve a detailed treatment.
In this article,
using basic differential geometry, we generalize
this problem to an arbitrary surface, including the effects of friction, 
providing a general expression to determine under which conditions
the particle leaves the surface. An
explicit integral form for the velocity is given, which is analytically
integrable for some cases, and we find general criteria to determine the
critical velocity at which the particle immediately leaves the surface.
Five curves, a circle, ellipse, parabola, catenary and cycloid, are analyzed
in detail. Several interesting features appear, for instance,
in the absense of friction, a particle moving on a catenary leaves the
surface just before touching the floor, and in the case of the parabola,
the particle never leaves the surface, regardless of the friction.
A phase diagram that separates the conditions that lead to a particle
stopping in the surface to those that lead to a particle departuring
from the surface is also discussed.

\end{abstract}

\maketitle 

%
%
\section{Introduction} 

%

\begin{figure}[h]
\centering
  \includegraphics[width=6cm]{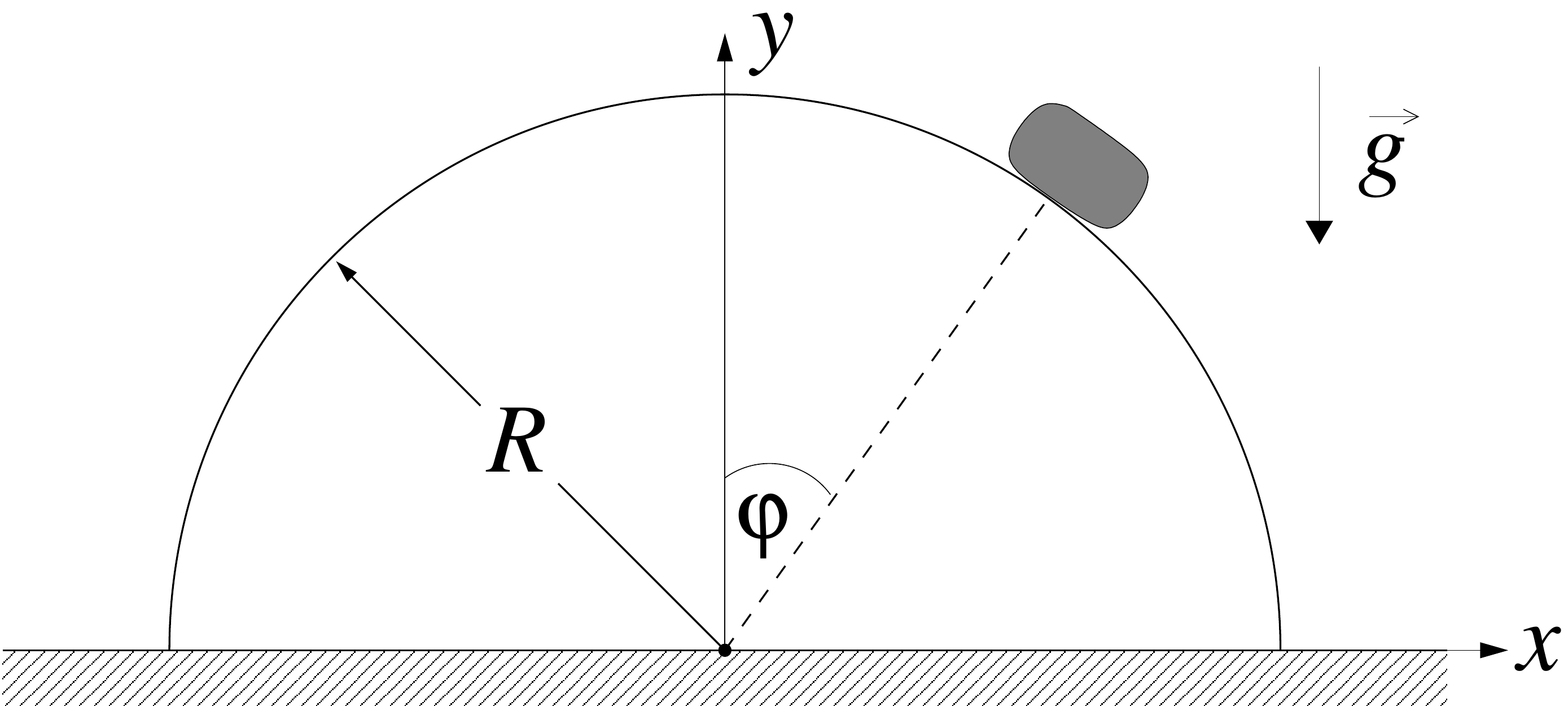}
  \caption{Particle sliding from the top of a semicircle under the action of gravity.}
  \label{f:af-8-25}
\end{figure}

A well studied problem in elementary mechanics, in one of its many
formulations, is the following~\cite[Chap. 8]{AF67}:
\textit{A boy/girl of mass $m$ is seated on a hemispherical mound of ice.
If he/she starts sliding from rest (assume the 
ice to be frictionless), where is the point P at which the boy/girl 
leaves the mound?} In terms of classical mechanics, it corresponds to the
motion of a point particle slipping from the top of a semicircle under
the action of gravity, as shown Fig.~\ref{f:af-8-25}.

It is difficult to know what is the original version of this problem,
but we can suppose that it comes after the influential books of
Francis Weston Sears~\cite{Ho99}. 
%
What is interesting about the problem is that it combines many topics. The usual
solution combines conservation of mechanical energy with elements from circular
motion (radial and tangential acceleration), together with the key condition
to the problem: the release point takes place when the normal force is zero.
The solution is independent of both the particle mass $m$ and the gravitational
acceleration $g$, resulting in the particle leaving the circular surface
when $\cos\phi=2/3$. This release point does not depend on the initial height
either, and its value is $\phi=\arccos(2/3)\approx48.19^{\circ}$.

A natural question that arises from studying this situation is how to
incorporate friction on the circular surface. In this case, the energy
is no longer conserved and it has been already solved in previous works~\cite{Mu03,PM07,LPM08}.
A more general version of the problem is to study
the point of departure in an arbitrary, frictionless
surface, described by a function $y=f(x)$. Here, the conservation of energy
does apply, and allows to write an equation in terms of $f$ and its
derivatives, whose solution gives the point of departure~\cite{Ag12}.
Although the treatment is general, it assumes that
the $y$ coordinate can be expressed as an explicit function of $x$, which is not
always possible (like a cycloid or a general conic section), and is also
limited to the cases where the mechanical energy is conserved  (no friction).
The case of rough surfaces (with friction) is only briefly
discussed for the cases in which $f(x)=-\alpha x^k$, without giving
an equation to determine the point of departure, but providing some
criteria to determine if the particle leaves the surface or not.

Our goal is to solve the following general problem: to find the point of departure of a
particle sliding in an arbitrary surface with friction, providing an expression
for the velocity of the particle, determining under which conditions the particle
stops on surface, and when it always remains on the surface. We also determine
the maximum initial velocity, above which the particle leaves the surface
immediately, and identify for which curves this maximum velocity is zero, i.e.,
the particle leaves the surface immediately, regardless of the initial velocity
given.
%

We pose the problem in terms of basic differential
geometry~\cite{doC76,Shifrin2015}, which gives us
a systematic way to deal with curves and allows us to obtain the velocity
of the particle as a general explicit expression. We consider some particular
famous curves, like parabola and ellipse, and later we provide a
classification of the curves that show the same behaviour.

%
%
\section{Theoretical Framework}
In this section, we review some basic concepts of differential geometry and present
the equations of motion of a particle sliding over an arbitrary surface written
in the Frenet--Serret frame, which are used to derive an expression for the
velocity of the particle at any point in the surface.

\subsection{Kinematics}
Consider a particle moving on the XY plane, describing an arbitrary trajectory.
Its position at any instant of time $t$ is given by
$\v r(t)=(x(t),y(t))$, the velocity by $\v v(t)=d\v r/dt$, and its acceleration
by $\v a=d\v v/dt$. The trayectory
is the collection of all points $\v r(t)$, which means that for a time interval
$[t_0,T]$, the function $\v r\colon[t_0,T]\to\mathbb R^2$ is a parametrized curve
whose image corresponds to the trajectory of the particle.
Conversely, if the particle is sliding on a surface, a parametrization of the
surface describes its trayectory along the surface.

\begin{figure}[h]\centering
  \includegraphics[width=8cm]{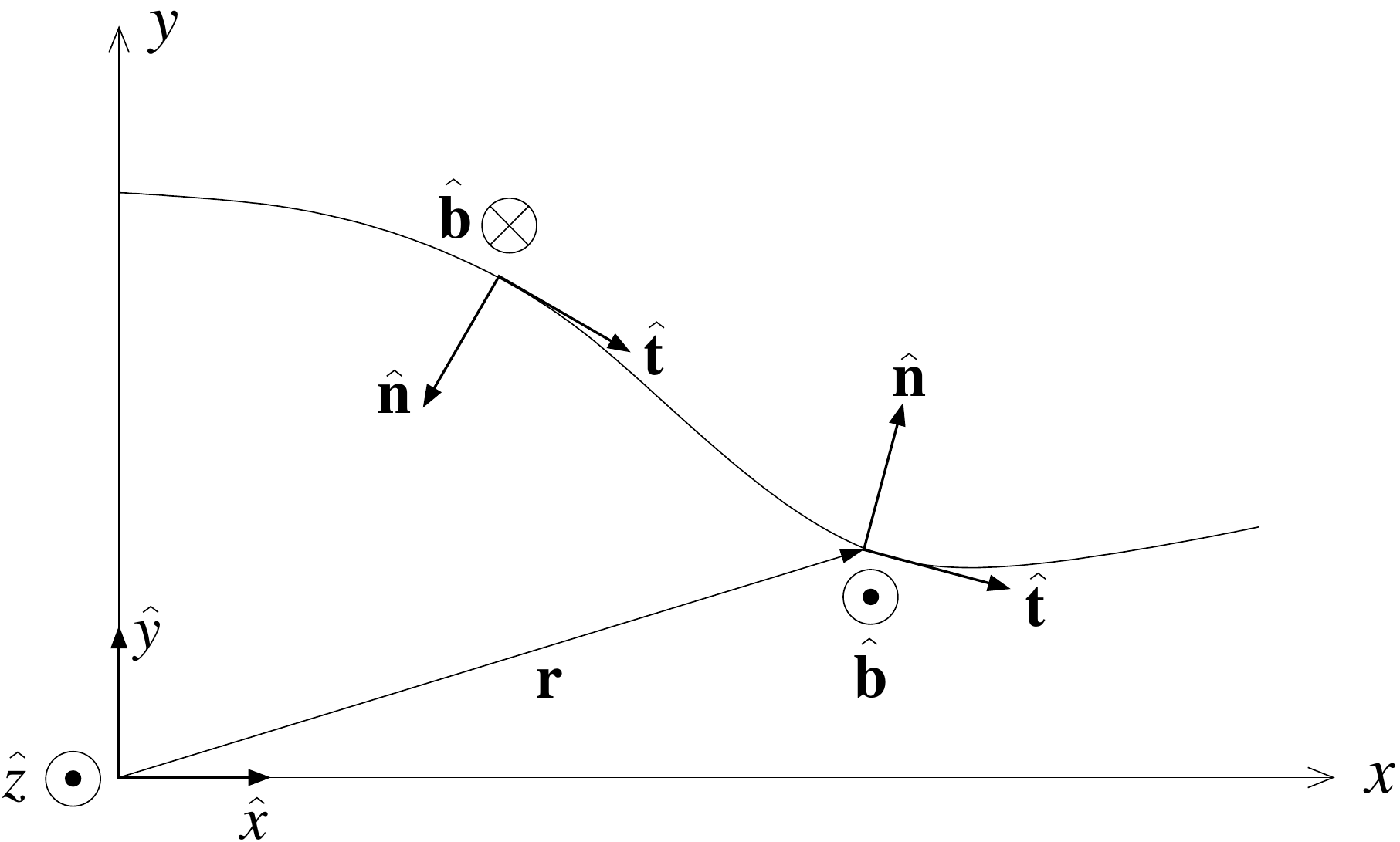}
  \caption{Tangent $\tun$, normal $\nun$, and binormal $\bun$ vectors of the Frenet--Serret frame, viewed from the Cartesian system.}
  \label{fig:Frame}
\end{figure}

A curve $\v r$ can always be reparametrized by its arclength 
\begin{equation}\label{eq:s(t)}
  s(t) = \int_{t_{0}}^{t} \left\|\frac{d\v r(\tau)}{d\tau}\right\| \, d\tau,
  \quad \text{where } t \in [t_{0}, T],
\end{equation}
as long as its derivative $d\v r(t)/dt\neq0$ in the interval, since
$ds(t)/dt=\|r'(t)\|=v(t)>0$ for all $t$ implies that $s$ is an increasing diferentiable function
and, therefore, has a differentiable inverse function $t=t(s)$. The
arclength also allows us to define the so called
Frenet--Serret frame, which is a coordinate system (orthonormal basis)
formed by three orthonormal vectors associated to any regular curve:
the tangent $\tun$, the normal $\nun$ and binormal $\bun$. They are defined by
\begin{subequations}\label{eq:frenet}
\begin{align}
\tun &\equiv \frac{d\v r}{ds}\\
\nun &\equiv \frac{1}{\kappa(s)}\frac{d\tun}{ds}\\
\bun &\equiv \tun\times\nun,
\end{align}
\end{subequations}
where $\kappa(s)\equiv\left\|d\tun/ds\right\|$ is the curvature of the curve at $s$.
Notice that the normal vector $\nun$ is not defined when $\kappa(s)=0$, which is the
case of flat surfaces and inflection points of functions.
The vector $\bun$ is out of the plane XY, and is parallel to the
unit vector $\zun$ of the Cartesian coordinates. Thus, $\bun\cdot\zun=\pm1$,
depending on the concavity of the curve (see Fig.~\ref{fig:Frame}).

%
%
The arclength
$s$ corresponds to the distance that the particle has travelled along the
surface, and can be used to describe the velocity and acceleration of the
particle:
%
%
%
\begin{eqnarray}
  \v v(t(s))&=& \frac{d\v r}{ds}\frac{ds}{dt}=v(t(s))\tun(s)\\
  \v a(t(s)) &=& \frac{dv(t(s))}{ds} \frac{ds}{dt} \tun(s)
  +v(t(s)) \frac{d\tun(s)}{ds} \frac{ds}{dt} .
\end{eqnarray}
If we introduce the notation $v(s)\equiv v(t(s))$ and $a(s)=a(t(s))$,
these equations take the form
%
%
\begin{eqnarray}
  \v v(s) &=& v(s)\tun(s)\\
  \v a(s) &=& v(s) \frac{dv(s)}{ds} \tun(s) + v^{2}(s) \kappa(s) \nun(s).
\end{eqnarray}

\subsection{Dynamics in the Frenet--Serret Frame}

The forces acting over a particle of mass $m$ sliding on a surface are its weight
$\v W=m\v g=m\left((\v g\cdot\v t)\tun+(\v g\cdot\v n)\nun\right)$, the normal
force $\v N$ and the friction $\v{f_r}=-\mu N\tun$,  where $\mu$ is the friction
coefficient and $N=\|\v N\|$. The normal force is given by
$\v N=(\bun\cdot\zun)N\nun$,
resulting in $\v N$ being parallel or antiparallel to $\nun$.

Using the Newton's laws,
$\v F=m\v a=\v W+\v{f_r}+\v N$ leads to
\begin{align}
  m v(s) \frac{dv(s)}{d s} &= m\v g \cdot \tun(s) - \mu N(s) \\\label{eq:Normal1}
  m \kappa(s) v^{2}(s) &= m\v g \cdot \nun(s) + (\bun\cdot\zun)N(s) .
\end{align}
%
Eliminating $N(s)$ from the equations, we obtain a differential equation for $v^2(s)$:
\begin{equation}
\label{e:ecdifv2}
 \frac12\frac{d}{ds} \left( v^{2}(s) \right) + 
  \frac{\mu \kappa(s)}{\bun\cdot\zun}\left(v^{2}(s)\right) = \ngr{g} \cdot \tun(s) + \mu \frac{\v g\cdot \nun(s)}{\bun\cdot\zun} .
\end{equation}
%
Multiplying by the integrating factor
$\exp\left(\frac{2\mu}{\bun\cdot\zun}\int_0^s\kappa(s')ds'\right)$,
the above equation can be rewritten as
\begin{equation}\label{eq:dv2ds}
 \frac{d}{ds}\left[  v^{2}(s) \e^{\frac{2\mu}{\bun\cdot\zun}\Phi(s)}\right]
 =2\e^{\frac{2\mu}{\bun\cdot\zun}\Phi(s)}\left[\v g\cdot\tun(s) + \mu \frac{\v g \cdot \nun(s)}{\bun\cdot\zun}\right],
\end{equation}
where we have defined
\begin{equation}\label{eq:Phi}
\Phi(s)\equiv\int_0^s \kappa(s')ds'.
\end{equation}
Integrating from 0 to $s$ at both sides of Eq.~\eqref{eq:dv2ds}, we obtain
\begin{eqnarray}\nonumber
  v^{2}(s) &=& v_0^2\e^{-\frac{2 \mu}{\bun\cdot\zun}\Phi(s)}
  +2 \e^{-\frac{2 \mu}{\bun\cdot\zun}\Phi(s)}\\\label{eq:velocity}
  &\times&\int_0^s \left[\v g\cdot \tun(s') + \mu \frac{\v g \cdot \nun(s')}{\bun\cdot\zun} \right] 
  \e^{\frac{2 \mu }{\bun\cdot \zun}\Phi(s')} ds'. 
\end{eqnarray}
where we have used that $\Phi(0)=0$ and $v(0)\equiv v_0$. This is the most
general equation for the velocity, since $s$ is always defined, increasing,
and unique. However, an anallytical expression for the arclength $s$ is
impossible to obtain in most of the particular cases, since the integral
in Eq.~\eqref{eq:s(t)} cannot be solved.
We will overcome this problem using the tangential angle, described in the next
section, acknowledging its restrictions, but being sufficient for our purposes.

The magnitude of the normal force can be obtained from Eq.~\eqref{eq:Normal1}:
\begin{equation}\label{eq:Normal}
N(s)=mv^2(s)\frac{\kappa(s)}{\bun\cdot\zun}-m\frac{\v g\cdot\nun(s)}{\bun\cdot\zun},
\end{equation}
where $v(s)$ is given by Eq.~\eqref{eq:velocity}. 
The release condition is $N(s)=0$, which leads to
\begin{equation}\label{eq:release}
\frac{v^2(s)}{gR}=\frac1{gR}\frac{\v g\cdot\nun(s)}{\kappa(s)}
\end{equation}
for some $s$, where $R$ is the height from which the particle starts moving.
Both sides of the above equation are dimensionless, and we will call the right-hand
side of this equation the {\it horizon} of the curve:
\begin{equation}\label{eq:horizon}
H(s)\equiv\frac1{gR}\frac{\v g\cdot\nun(s)}{\kappa(s)}.
\end{equation}
The particle remains in the surface as long as the normal is greater than zero.
For surfaces that are concave downwards (such as the left portion of the curve in
Fig.~\ref{fig:Frame}), $\bun\cdot\zun=-1$, and the condition $N(s)\geqslant0$
becomes, using Eq.~\eqref{eq:Normal},
\begin{equation}
\frac{v^2(s)}{gR}\leqslant H(s).
\end{equation}
The horizon, then, represents the frontier that the velocity has to reach in
order to release the particle from the surface. 
In particular, for $s=0$ we have
$v_0^2/gR \leqslant H(0)=\v g\cdot\nun/(gR\kappa(0))$.
If both sides are equal ($v_0$ touches the horizon), the particle leaves the
surface immediately, flying in a parabolic trajectory above the surface.
Therefore, the maximum initial velocity that can be given to the particle is
given by
\begin{equation}\label{eq:maxinitvel}
\left(\frac{v_0^2}{gR}\right)^{(\text{max})}\equiv H(0)=\frac{\v g\cdot\nun(0)}{gR\kappa(0)}.
\end{equation}
When the surface is concave upwards (like the right portion of the curve in
Fig.~\ref{fig:Frame}), then $\bun\cdot\zun=1$, and the remain-on-the-surface
condition, $N(s)\geqslant0$, implies that $v^2(s)/gR\geqslant H(s)$. Therefore
the initial velocity must satisfy $v_0^2/gR\geqslant H(0)$.
Since $\v g\cdot\nun(0)<0$ in this case, we have $H(0)\leqslant0$, therefore
$v_0^2/gR\geqslant H(0)$ for all concave-upwards surfaces. This means that the
initial velocity has no restrictions (the particle always stays in the surface
if the surface is concave upwards).


\subsection{Tangential Angle}
We are interested in ``hill-down'' curves (concave downwards),
which means that the vertical component
of the position decreases monotonically. For these curves, $\tun$ is a vector
with a positive $x$ component and negative $y$ component. Then, a useful way
of decribing the tangent vector is through the Whewell angle~\cite{doC76,Shifrin2015,Wh49}
\begin{equation}
\tun = \cos\phi\xun-\sin\phi\yun,
\end{equation}
where $\phi$ is the angle that $\tun$ forms with the horizontal
(the same that $\nun$ forms with the vertical, see Fig.~\ref{fig:circle}).
With this definition, the normal becomes
\begin{equation}
\nun = -\frac{(d\phi/ds)}{\left|d\phi/ds\right|}[\sin\phi\xun+\cos\phi\yun].
\end{equation}
The curvature is defined by $\kappa(s)=\|d\tun/ds\|$, therefore
$
\kappa(s)=|d\phi/ds|.
$
If $\phi$ is monotonically increasing (the curve is concave downwards),
$d\phi/ds>0$, therefore $\kappa(s)=d\phi/ds$.
This means that $\int \kappa(s)ds=\phi(s)$, and therefore $\Phi$ in Eq.~\eqref{eq:Phi}
becomes $\Phi(s)=\phi(s)-\phi_0.$
We also have $\bun=-\zun$, then $\bun\cdot\zun=-1$.
This allows us to write Eq.~\eqref{eq:velocity} as
\begin{eqnarray}\nonumber
  v^{2}(\phi) &=& v_0^2\e^{2 \mu(\phi-\phi_0)} +2 g\e^{2\mu\phi}\\\label{eq:velocityphi-inc}
  &&\times\int_{\phi_0}^\phi (\sin\phi' - \mu \cos\phi') 
  \frac{\e^{-2 \mu \phi'}}{\kappa(\phi')}\,d\phi',
\end{eqnarray}
where we have used $\v g=-g\yun$. The horizon in Eq.~\eqref{eq:horizon} can also be
written in terms of $\phi$:
\begin{equation}\label{eq:horizon-phi}
H(\phi)=\frac1{gR}\frac{\v g\cdot\nun}{\kappa(\phi)}=\frac{\cos\phi}{R\kappa(\phi)}.
\end{equation}
When $\phi$ is a decreasing function (the curve is concave upwards), $\kappa=-d\phi/ds$ and
\mbox{$(\bun\cdot\zun)$}$=1$,
then we have
\begin{eqnarray}\nonumber
  v^{2}(\phi) &=& v_0^2\e^{2 \mu(\phi-\phi_0)} -2 g\e^{2\mu\phi}\\\label{eq:velocityphi-dec}
  &&\times\int_{\phi_0}^\phi (\sin\phi' - \mu \cos\phi') 
  \frac{\e^{-2 \mu \phi'}}{\kappa(\phi')}\,d\phi'.
\end{eqnarray}
In the following section, we will only analyze curves which are concave
downwards, since we are interested in the point of departure of a particle
slipping on its surface, which does not appear in concave upwards curves.


%

%
%
\section{Results}
We will analyze some particular curves in the framework presented,
characterizing the movement of a particle falling on their surface.
In all cases, the particle starts its motion from the top of a curve,
at a position $\v r(0)=R\yun$.
\subsection{Departure from the Circle}\label{sec:circle}

\begin{figure}[!h]
\includegraphics[width=6cm]{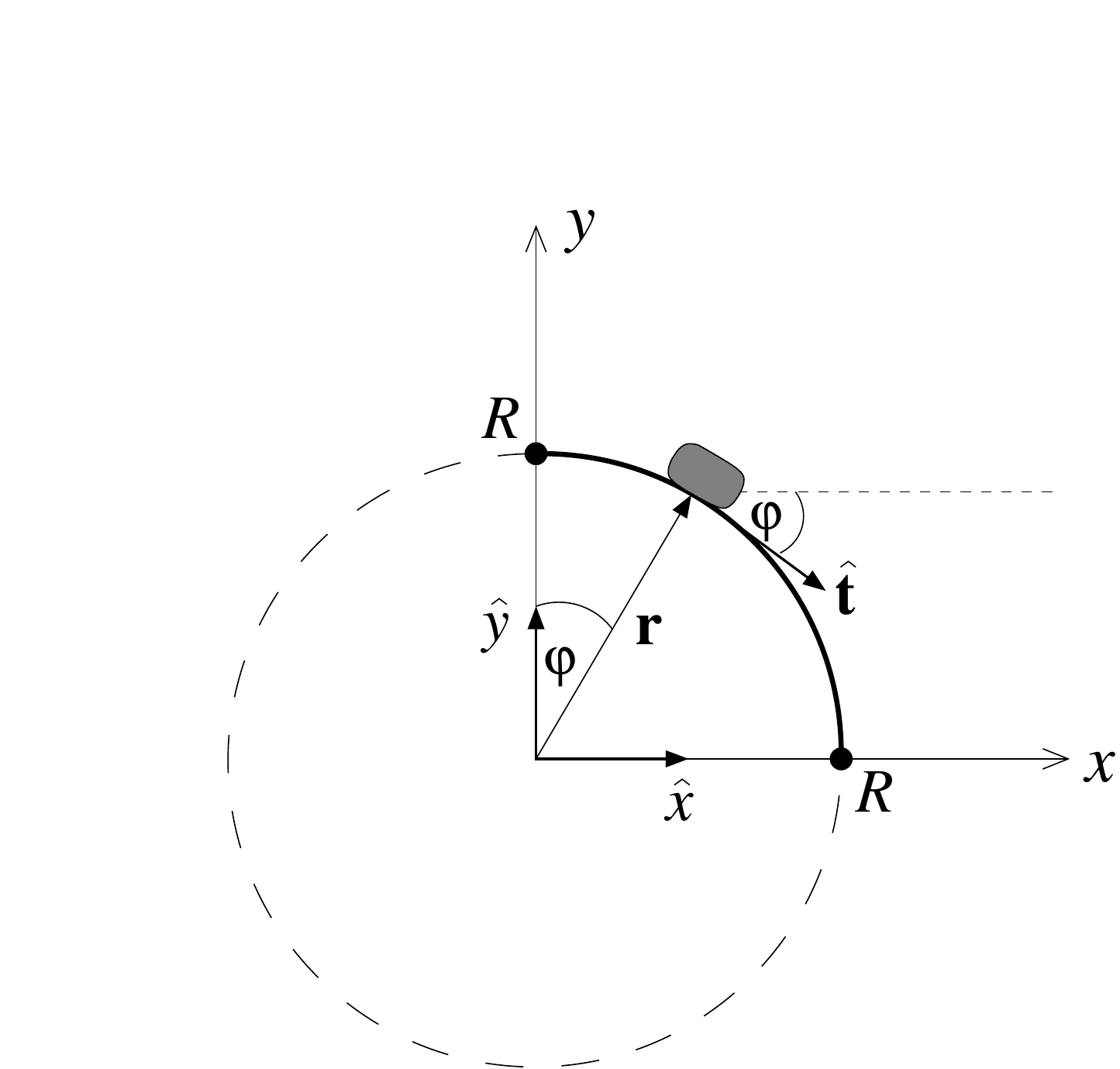}
\caption{A circle parametrized by $\v r(\phi) = ( R \sin\phi , R \cos\phi)$.} 
\label{fig:circle}
\end{figure}

We can parametrize the circle using the tangential angle $\phi$ by
$\v r(\phi) = ( R \sin\phi , R \cos\phi)$,
where $\phi \in [0, \pi/2]$ also corresponds to the angle between
$\v r(\phi)$ and the vertical axis, as shown in Fig.~\ref{fig:circle}.
Note that the circle is the only curve in which the tangential angle $\phi$
coincides with the angle that the vector $\v r$ forms with the vertical, which
is not the case of other curves, as we will see in the following sections.
For the circle, we have $\kappa(\phi)=1/R$, and since $\phi$ is
increasing, Eq.~\eqref{eq:velocityphi-inc} leads to
\begin{equation}\label{eq:velocitycircle}
  \frac{v^{2}(\phi)}{gR} = \frac{v_0^2}{gR}\e^{2 \mu\phi} +2\e^{2\mu\phi}
  \int_{0}^\phi (\sin\phi' - \mu \cos\phi')\e^{-2 \mu \phi'}\,d\phi'.
\end{equation}
This integral can be easily solved by replacing
$\sin\phi=(\e^{i\phi}-\e^{-i\phi})/2i$ and
$\cos\phi=(\e^{i\phi}+\e^{-i\phi})/2$, leading to an explicit form for
the velocity:
\begin{equation}
\frac{v^2(\phi)}{gR}=\frac{v_0^2}{gR}\e^{2\mu\phi}+\frac{(2-4\mu^2)\left(e^{2\mu\phi}-\cos\phi\right)-6\mu\sin\phi}{1+4\mu^2}.
\end{equation}
This equation is identical to the one found by Mungan~\cite{Mu03}.

%
%
%

%
For the circle, the horizon in Eq.~\eqref{eq:horizon-phi} becomes $H(\phi)=\cos\phi$,
which means that, before leaving the surface, the velocity satisfies
\begin{equation}\frac{v^2(\phi)}{gR}\leqslant\cos\phi,\end{equation}
and, according to Eq.~\eqref{eq:maxinitvel}, the maximum initial velocity that can
be given to the particle is
\begin{equation}
\left(\frac{v_0^2}{gR}\right)^{(\text{max})}=1.
\end{equation}
For the case $\mu=0$, we recover the well-known expression found in text books:
$$v^2(\phi)=v_0^2+2gR(1-\cos\phi),$$
which can be directly obtained from conservation of energy.
In this case, the particle leaves the surface (or touches the horizon) when
$$\cos\phi=\frac23+\frac{v_0^2}{3gR},$$
recovering the already known departure angle for the frictionless circle
with $v_0=0$, $\phi=\arccos(2/3)\approx48.19^\text{o}$.

\begin{figure}[!h]
\includegraphics[width=7cm]{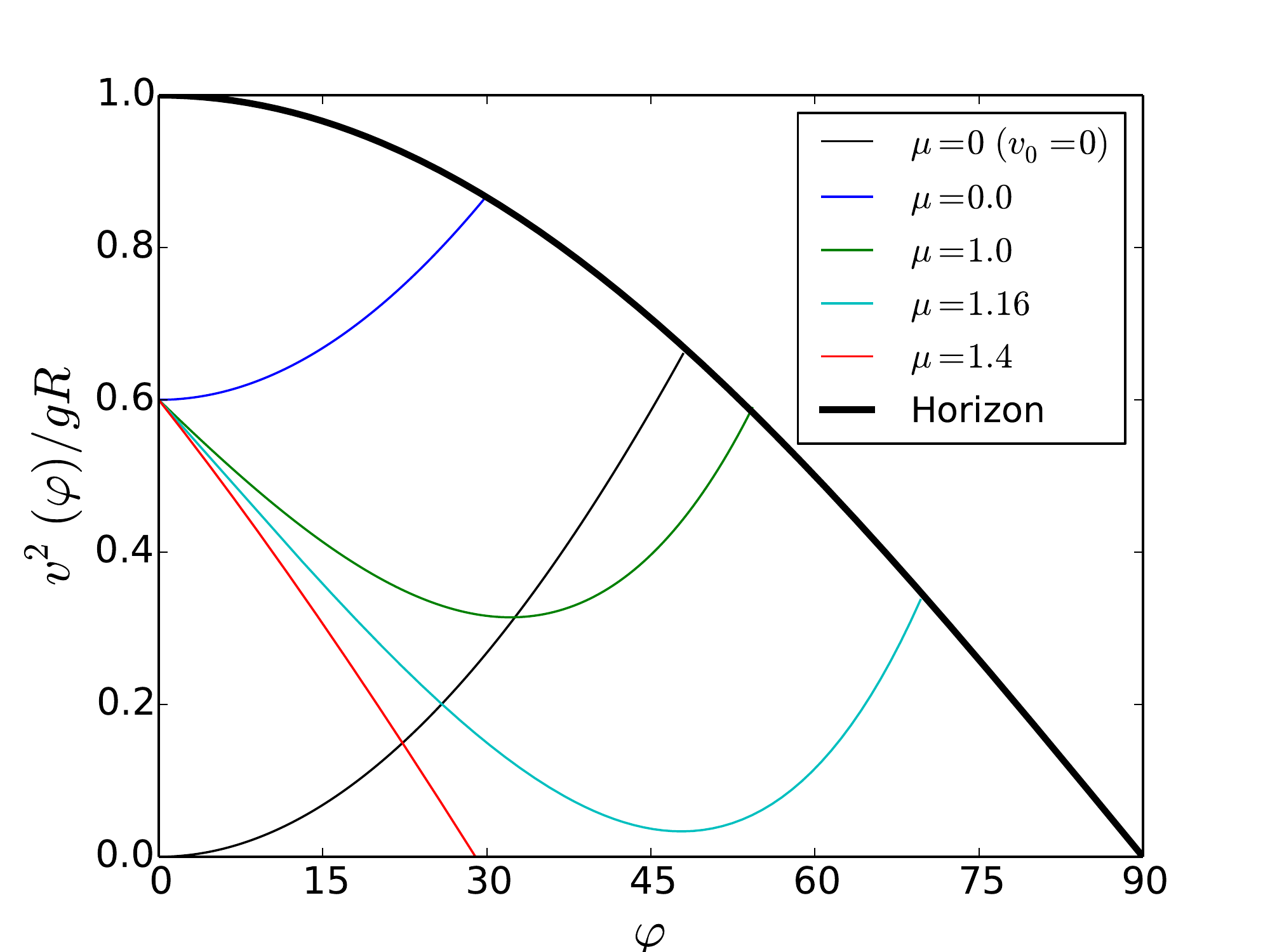}
\caption{Squared velocity as a function of $\phi$ and the horizon of the circle,
$H(\phi)=\cos\phi$ for $v_0^2/gR=0.6$. As pointed out by Mungan~\cite{Mu03},
the horizon is equal to $v^2(\phi)/gR$ for the case $v_0^2/gR\to1$ and
$\mu\to\infty$.}
\label{fig:horizon-circle}
\end{figure}

In Fig.~\ref{fig:horizon-circle} we show the evolution of $v^2(\phi)/gR$ as it
approaches the horizon $H(\phi)=\cos\phi$.
The thin, black line corresponds to the velocity of a particle slipping in a
frictionless, circular surface. We observe that the intersection with the
horizon (thick, black line) takes place at the already predicted angle
$\phi\approx48.19^\text{o}$, which is the point at which the particle leaves
the surface. The evolution of the velocity is also shown for different values
of $\mu$ for an initial velocity of $v_0^2/gR=0.6$: with no friction
(blue line), $\mu=1.0$ (green) and $\mu=1.16$ (cyan). In all these cases, we
observe that the friction decreases the velocity up to certain point, in which
the gravity overcomes friction and the velocity starts increasing
again, until it reaches the horizon (the point at which the particle leaves
the surface). Notice that the more we increase the friction coefficient for
a given initial velocity, the farther the particle travels on the surface,
departuring at even greater angles.

However, if the friction coefficient is too large, it leads to solutions in
which the particle comes to rest in the surface
(red line Fig.~\ref{fig:horizon-circle}). This phenomenon suggests that, for
a given curve, there must be a critical friction coefficient for every initial
velocity (or a critical velocity for every friction coefficient)
such that the particle will get stuck on the surface for any value of $\mu$
greater than this critical coefficient. Conversely, for a given friction
coefficient, there must be a critical initial velocity such that, for initial
velocities lower than the critical one, the particle will always get stuck
on the surface.
We will analyze this phenomenon in detail in section~\ref{sec:phase-separation}.

\subsection{Departure from an Ellipse}

\begin{figure}[!h]\centering
\includegraphics[width=6cm]{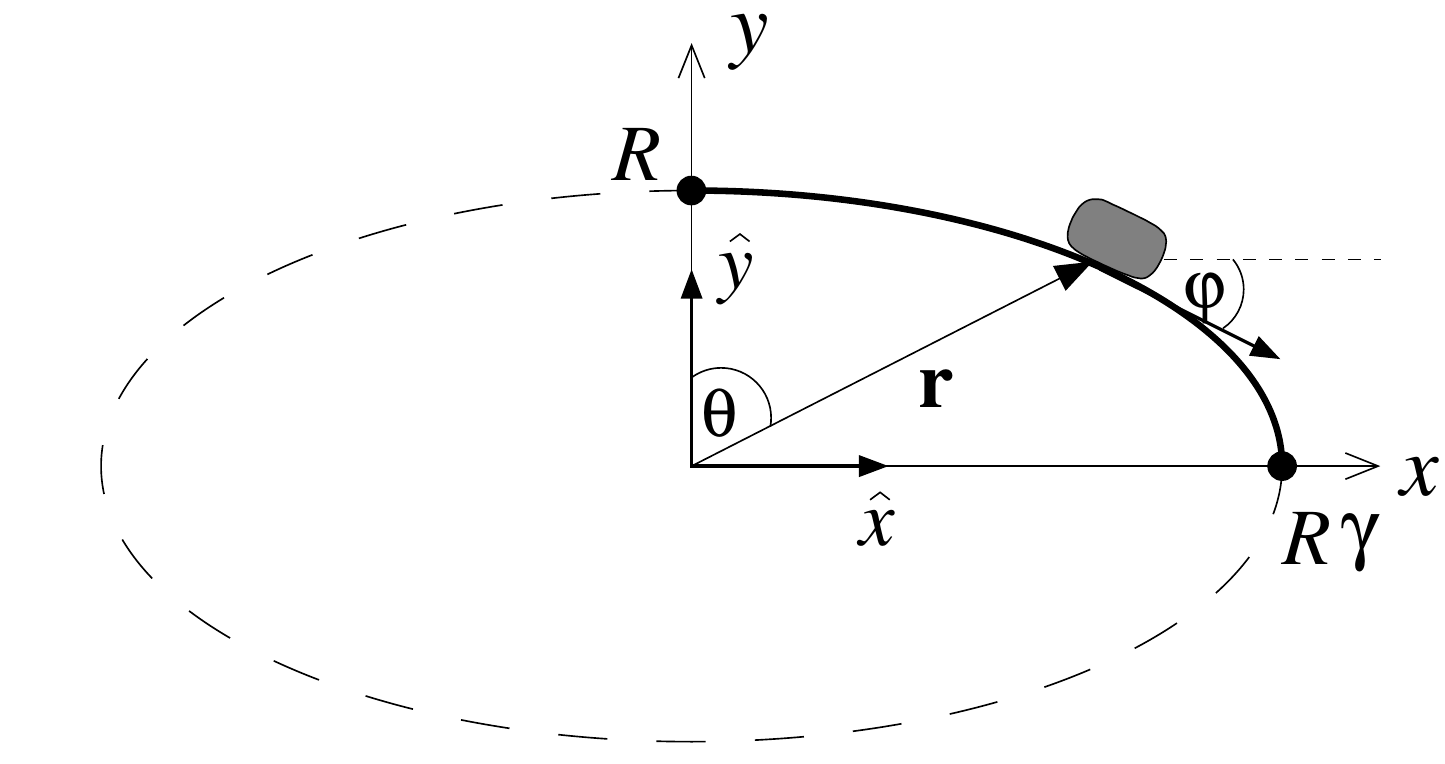}
\caption{An ellipse represented by $\v r(\alpha)=(R\gamma\sin\alpha,R\cos\alpha)$.
Notice that $\phi$ is not equal to $\theta$, the angle that $\v r$ forms with
the vertical, and both of them are different from $\alpha$. They are related by
$\tan\theta=\gamma\tan\alpha=\gamma^2\tan\phi$.}
\label{fig:elipse}
\end{figure}

We can parametrize the ellipse by
\mbox{$\v r(\alpha)=(R\gamma\sin\alpha,R\cos\alpha)$},
with $\alpha\in[0, \pi/2]$ and $\gamma>0$, which represents the curve shown in
Fig.~\ref{fig:elipse}.
The tangent vector in this case is
$$\tun=\frac{(R\gamma\cos\alpha\xun-R\sin\alpha\yun)}
{\sqrt{R^2\gamma^2\cos^2\alpha+R^2\sin^2\alpha}},$$
and since $\tun=\cos\phi\xun-\sin\phi\yun$, we obtain
the relationship
\begin{equation}\label{eq:phi-alpha-ellipse}
\tan\phi=\frac1\gamma\tan\alpha.
\end{equation}
This allows us to calculate the curvature in terms of $\phi$:
\begin{equation}
\kappa(\phi)=\frac{(\gamma^2\sin^2\phi+\cos^2\phi)^\frac32}{\gamma^2R}.
\end{equation}
Then, according to Eq.~\eqref{eq:velocityphi-inc}, the velocity becomes
\begin{eqnarray}\nonumber
\frac{v^2(\phi)}{gR}&=&\frac{v_0^2}{gR}\e^{2\mu\phi}+2\e^{2\mu\phi}\\\label{eq:velocityellipse}
&\times&\int_0^\phi\frac{\gamma^2[\sin\phi'-\mu\cos\phi']}{(\gamma^2\sin^2\phi+\cos^2\phi)^\frac32}\e^{-2\mu\phi'}d\phi',
\end{eqnarray}
which is not integrable analitically. However, we can still determine the
maximum initial velocity like we did in the last section: the horizon of the
ellipse is
\begin{equation}
H(\phi)=\frac{\gamma^2\cos\phi}{(\gamma^2\sin^2\phi+\cos^2\phi)^{3/2}},
\end{equation}
and its value in $\phi=0$ gives us the maximum initial velocity:
\begin{equation}
\left(\frac{v_0^2}{gR}\right)^{(\text{max})}=\gamma^2.
\end{equation}
For $\mu=0$, the velocity takes the form
\begin{equation}
\frac{v^2(\phi)}{gR}=
\frac{v_0^2}{gR}+2-\frac{2\cos\phi}{\sqrt{\gamma^2\sin^2\phi+\cos^2\phi}},
\end{equation}
which, in terms of $\alpha$, becomes
\begin{equation}
\frac{v^2(\alpha)}{gR}=\frac{v_0^2}{gR}+2(1-\cos\alpha).
\end{equation}
When this velocity reaches the horizon, which in terms of $\alpha$ becomes
$H(\alpha)=\cos\alpha(\gamma^2\cos^2\alpha+\sin^2\alpha)$, we obtain
\begin{equation}
\frac{v_0^2}{gR}+2=3\cos\alpha+(\gamma^2-1)\cos^3\alpha,
\end{equation}
which give us the departure angle $\alpha$. We can then obtain it in terms of
$\phi$ using Eq.~\eqref{eq:phi-alpha-ellipse}. This equation always has a
solution, as long as the condition $v_0^2/gR\leq\gamma^2$
($v_0\leq v_0^{(\text{max})}$). The solution also reobtains the departure
angle for the circle setting $\gamma=1$, where $\alpha=\phi$.

\begin{figure}[!h]
\includegraphics[width=7cm]{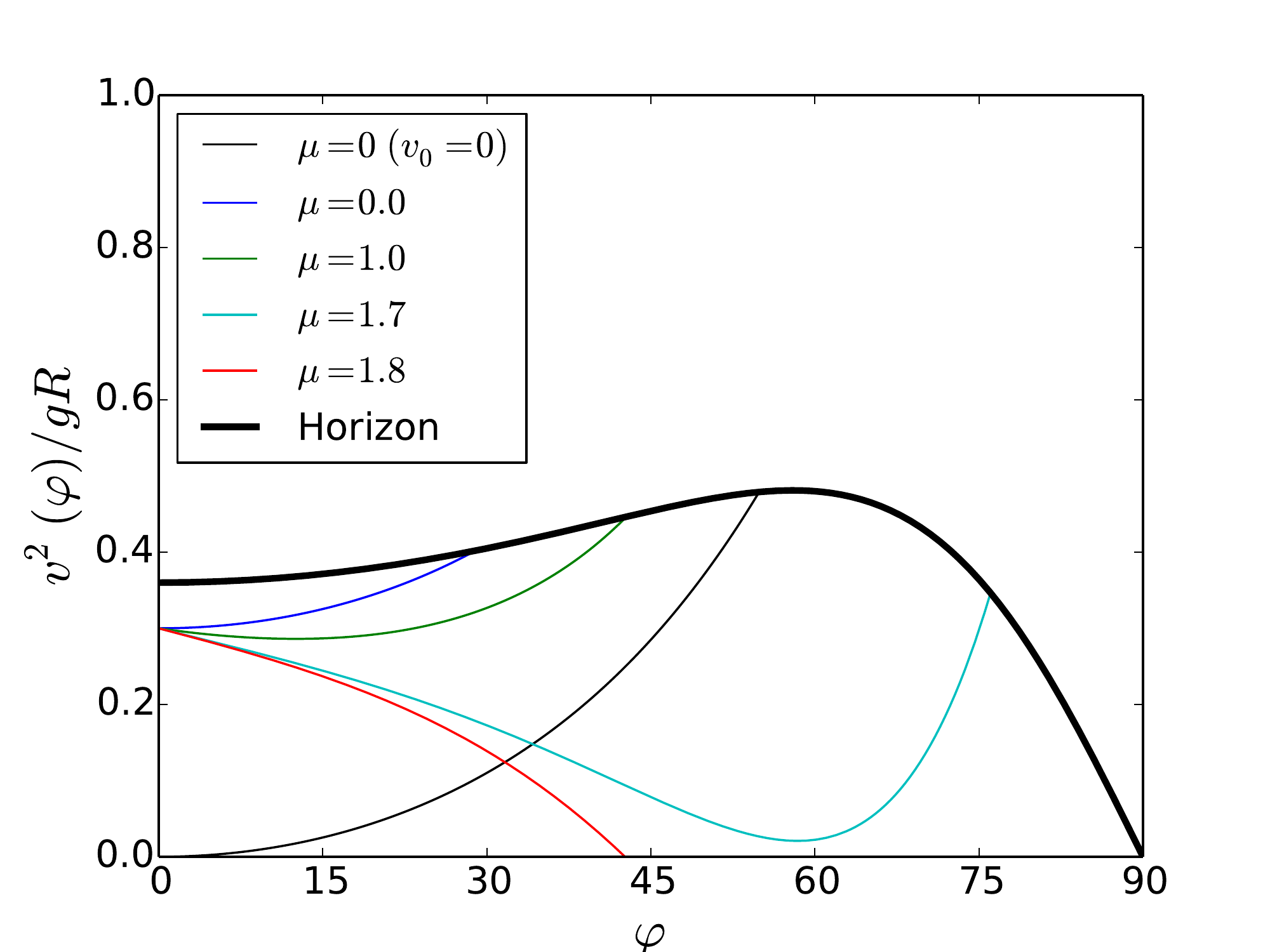}
\caption{Squared velocity as a function of $\phi$ and the horizon of the 
ellipse, $H(\phi)=\gamma^2\cos\phi/(\gamma^2\sin^2\phi+\cos^2\phi)^{3/2}$
 for $\gamma=0.6$, $v_0^2/gR=0.3$.} 
\label{fig:horizon-ellipse}
\end{figure}

In Fig.~\ref{fig:horizon-ellipse} we show the evolution of $v^2(\phi)/gR$ as it
approaches the horizon $H(\phi)=\gamma^2\cos\phi/(\gamma^2\sin^2\phi+\cos^2\phi)^{3/2}$.

\subsection{Departure from a Parabola}

\begin{figure}[!h]\centering
\includegraphics[width=6cm]{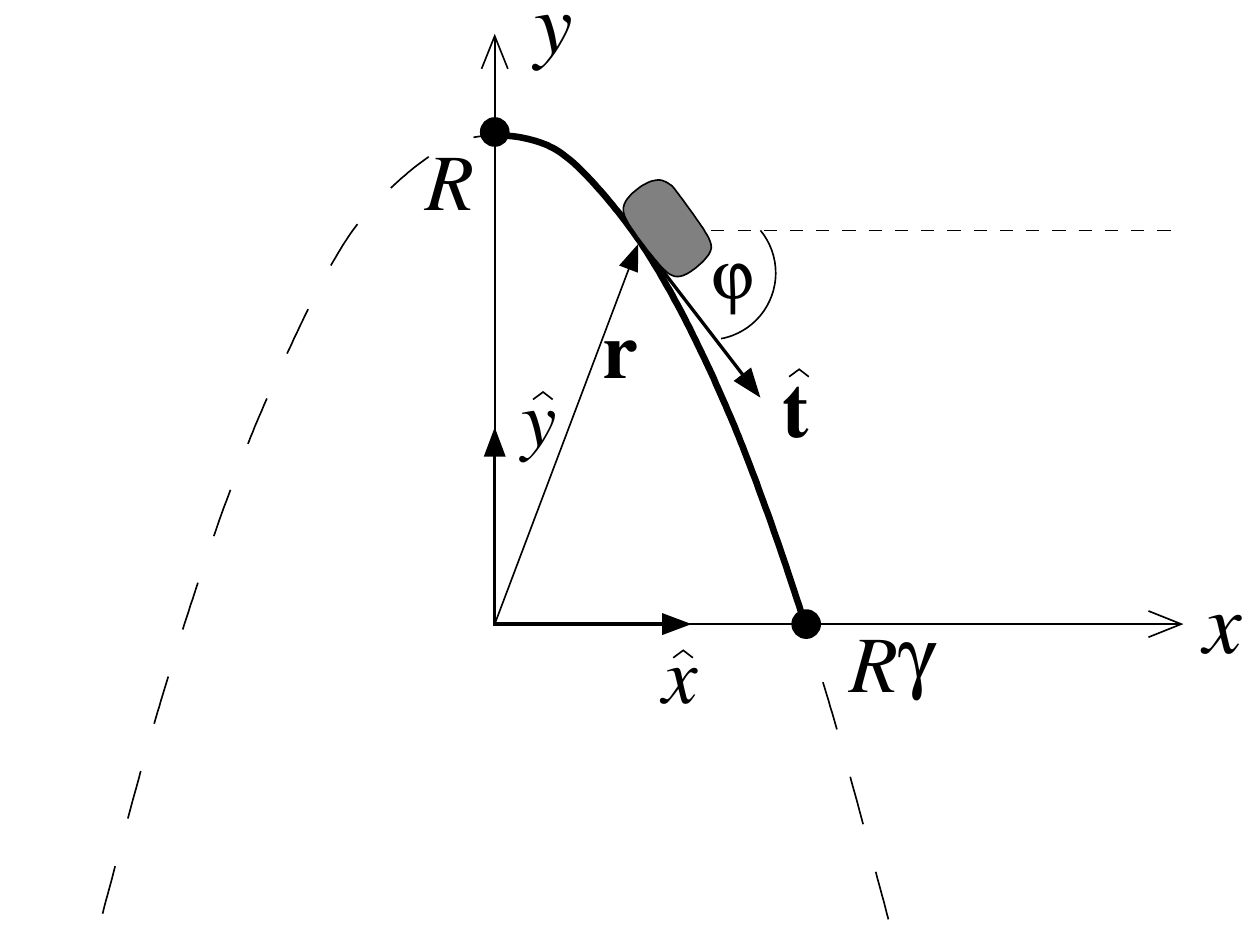}
\caption{A parabola represented by $\v r(\alpha)=(R\gamma\alpha,R(1-\alpha^2))$.}
\label{fig:parabola}
\end{figure}

Using the parametrization
$\v r(\alpha)=(R\gamma\alpha,R(1-\alpha^2))$
with $\alpha\in[0,\infty)$ and $\gamma>0$, we can describe the curve shown in
Fig.~\ref{fig:parabola}.
For this parametrization, we obtain a tangent vector
$$\tun =\frac{(\gamma\xun-2\alpha\yun)}{\sqrt{\gamma^2+4\alpha^2}},$$
which is equal to $\tun=\cos\phi\xun-\sin\phi\yun$, whereof the relationship
\begin{equation}
\tan\phi=2\alpha/\gamma
\end{equation}
is obtained. From this, the curvature in tems of $\phi$ is
\begin{equation}\label{eq:curvature-parabola}
\kappa(\phi)=\frac{2}{\gamma^2R}\cos^3\phi.
\end{equation}
Then, according to Eq.~\eqref{eq:velocityphi-inc}, the velocity becomes
$$
\frac{v^2(\phi)}{gR}=\frac{v_0^2}{gR}\e^{2\mu\phi}+\gamma^2\e^{2\mu\phi}
\int_0^\phi[\sin\phi'-\mu\cos\phi']\frac{\e^{-2\mu\phi'}}{\cos^3\phi'}d\phi'.
$$
An interesting fact is that
$$\frac{d}{d\phi}[\e^{-2\mu\phi}\sec^2\phi]=\frac{2\e^{-2\mu\phi}}{\cos^3\phi}[\sin\phi-\mu\cos\phi],$$
which makes the integral fully solvable, thus obtaining an explicit solution
for the velocity:
\begin{equation}\label{eq:velocityparabola}
\frac{v^2(\phi)}{gR}=\frac{\gamma^2}{2}\sec^2\phi
-\left(\frac{\gamma^2}{2}-\frac{v_0^2}{gR}\right)\e^{2\mu\phi}.
\end{equation}
We observe that the particular velocity $v_0^2/gR=\gamma^2/2$ anihilates all
dependence on the friction, leading to
\begin{equation}\label{eq:limitvelocityparabola}
\frac{v^2(\phi)}{gR}=\frac{\gamma^2}2\sec^2\phi
\end{equation}
for any value of $\mu$.
This is surprising, since no other curve shows this phenomenon.
The explanation is related to the fact that the trajectory of the particle
in free space under the action of gravity, is precisely a parabola. 

The horizon of the parabola is
\begin{equation}
H(\phi)=\frac{\gamma^2}{2}\sec^2\phi
\end{equation}
and the limit for the initial velocity in this case is
\begin{equation}
\left(\frac{v_0^2}{gR}\right)^{(\text{max})}=\frac{\gamma^2}2.
\end{equation}
Then, the particular value for which the velocity no longer depends on the
friction is actually the maximum initial velocity.
If the particle is given an initial velocity
$v_0\geq v_0^{(\text{max})}$, it will fly on a trayectory given by
$$x(t)=v_0t,\quad y(t)=R-\frac12gt^2.$$
Replacing $t=x/v_0$ for $y(t)$, we obtain
\begin{equation}
y=R-\frac{g}{2}\frac{x^2}{v_0^2}=R\left(1-\frac{x^2}{2R^2(v_0^2/gR)}\right),
\end{equation}
In our parametrization, $x=R\gamma\alpha$ and $y=R(1-\alpha^2)$, then
\begin{equation}
y=R(1-\alpha^2)=R\left(1-\frac{x^2}{2R^2(\gamma^2/2)}\right).
\end{equation}
Then, we can clearly see that as the velocity $v_0^2/gR$, given to the particle
to leave the surface, approaches to $\gamma^2/2$, the maximum velocity allowed
in the surface, the two parabolic trajectories become equal. This means that
the particle flies over the surface at all times, which is the reason why this
velocity eliminates the dependence on $\mu$ in Eq.~\eqref{eq:velocityparabola}.

The second interesting fact in this case is that when the second term
of Eq.~\eqref{eq:velocityparabola} is positive; that is, in the case that
$v_0^2/gR<\gamma^2/2$, we have
\begin{equation}
\frac{v^2(\phi)}{gR}<\frac{\gamma^2}{2}\sec^2\phi=H(\phi),
\end{equation}
for any value of $\mu$ and $v_0$. We have concluded, then, that the velocity
never reaches the horizon, which means that the particle never leaves the
parabolic surface, i.e., there is no departure angle.

For $\mu=0$, the velocity becomes
\begin{equation}\label{eq:velocityparabola-mu=0}
\frac{v^2(\phi)}{gR}=\frac{v_0^2}{gR}+\frac{\gamma^2}{2}\tan^2\phi
\end{equation}
and if $v_0=v_0^{(\text{max})}=\gamma\sqrt{gR/2}$, we recover 
$v^2(\phi)/gR=\frac12\gamma^2\sec^2\phi=H(\phi)$ obtained
in Eq.~\eqref{eq:limitvelocityparabola}. 

\begin{figure}[!h]
\includegraphics[width=7cm]{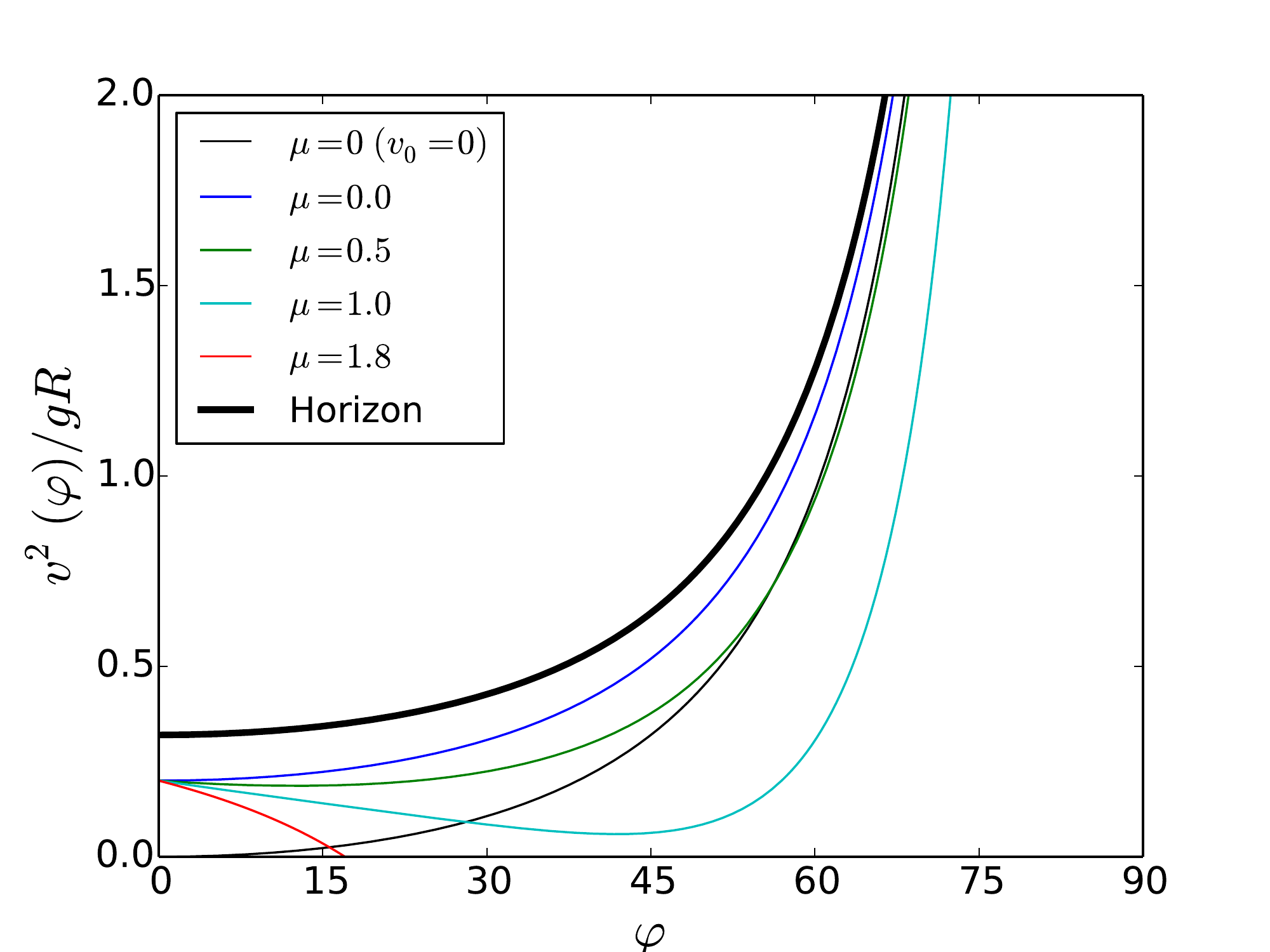}
\caption{Squared velocity as a function of $\phi$ and the horizon of the parabola.
$H(\phi)=\frac12\gamma^2\sec^2\phi$ for $\gamma=0.8$, $v_0^2/gR=0.2$.
For some conditions, the particle sticks to the surface, but when it does not
stop, it never touches the horizon (no departure from the surface).} 
\label{fig:horizon-parabola}
\end{figure}

In Fig.~\ref{fig:horizon-parabola} we show the evolution of $v^2(\phi)/gR$ as it
approaches the horizon $H(\phi)=\frac12\gamma^2\sec^2\phi$, never reaching it.

\subsection{Departure from a Catenary}

\begin{figure}[!h]\centering
\includegraphics[width=7cm]{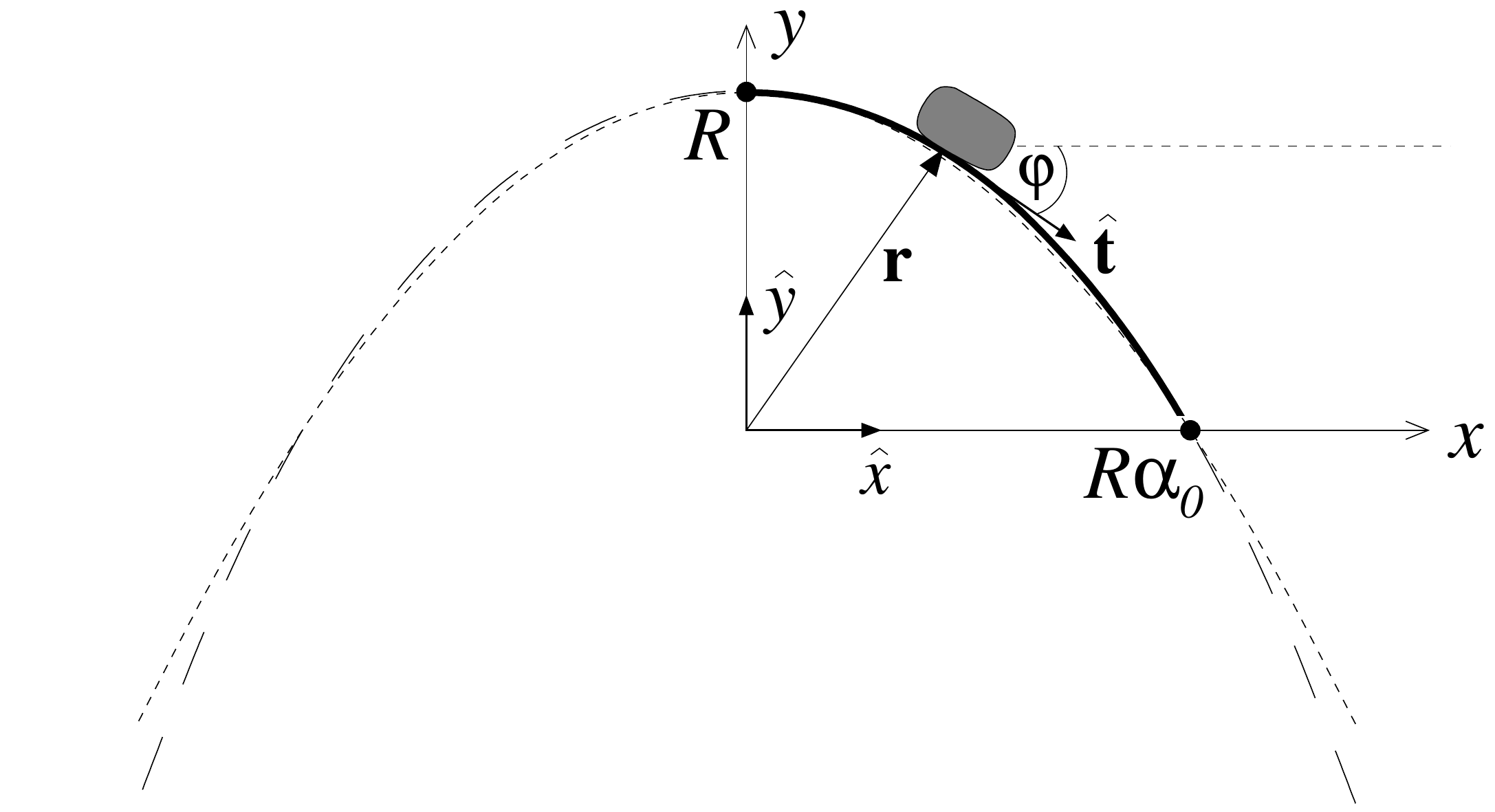}
\caption{A catenary represented by $\v r(\alpha)=(R\alpha,R(2-\cosh\alpha))$.
Here, $\alpha_0=\cosh^{-1}(2)\approx1.317$. The long-dashed line corresponds to
the catenary. The dotted line is a parabola (shown for reference) that crosses
through the points $(0,R)$ and $(R\alpha_0,0)$.}
\label{fig:catenaria}
\end{figure}

The parametrization $\v r(\alpha)=(R\alpha,R(2-\cosh\alpha))$
represents the catenary shown in Fig.~\ref{fig:catenaria}.
Under this parametrization,
$$\tun=\sech(\alpha)\xun-\tanh\alpha\yun,$$
which leads to
\begin{equation}
\tan\phi=\sinh\alpha.
\end{equation}
From this relationship, we can obtain the curvature in term of $\phi$:
\begin{equation}
\kappa(\phi)=\frac1R\cos^2\phi.
\end{equation}
Notice that, although the parabola that intersects the vertical and horizontal
axes in the same points as the catenary (see dotted line in
Fig.~\ref{fig:catenaria}), is very similar to the catenary, their curvatures
are not. From Eq.~\eqref{eq:curvature-parabola}, we obtain that the curvature of this
particular parabola is
$\kappa(\phi)=(2/\alpha_0^2R)\cos^3\phi\approx1.153\cos^3\phi$, while the
catenary has a curvature $\kappa(\phi)=\cos^2\phi/R$. Interestingly, both curves
have the same curvature when $\cos\phi=\alpha_0^2/2$
($\phi\approx29.86^\text{o}$).

If we replace the curvature of the catenary in Eq.~\eqref{eq:velocityphi-inc}, we
obtain the velocity of the particle slipping on this surface,
\begin{equation}
\frac{v^2(\phi)}{gR}=
\frac{v_0^2}{gR}\e^{2\mu\phi}+2\e^{2\mu\phi}
\int_0^\phi[\sin\phi'-\mu\cos\phi']\frac{\e^{-2\mu\phi'}}{\cos^2\phi'}d\phi'.
\end{equation}
%
%
%
The horizon is
\begin{equation}
H(\phi)=\sec\phi
\end{equation}
which bounds the initial velocity to
\begin{equation}
\left(\frac{v_0^2}{gR}\right)^{(\text{max})}=H(0)=1,
\end{equation}
the same as the circle, since $\kappa(0)=1/R$ in both cases.
For $\mu=0$, the velocity is
\begin{equation}
\frac{v^2(\phi)}{gR}=\frac{v_0^2}{gR}+2(\sec\phi-1),
\end{equation}
which touches the horizon when
$$
\frac{v_0^2}{gR}+2(\sec\phi-1)=\sec\phi,
$$
or
\begin{equation}\label{eq:intersec-catenary}
\sec\phi=2-\frac{v_0^2}{gR}.
\end{equation}
If the particle starts with zero velocity, it leaves the surface at a departure
angle $\phi_0$ that satisfies $\sec\phi_0=2$, that is, when
$\phi_0=60^{\text{o}}$. Therefore, the particle leaves the frictionless catenary
with a steeper slope than in the frictionless circle, when the particle starts
with zero velocity.
In terms of the parameter $\alpha$, this means that
the particle leaves the catenary  for a value $\alpha_0$ that satisfies
$\cos\alpha_0=2$ (since $\sec\phi=\cosh\alpha$). Therefore
the particle leaves the surface when
$\v r(\alpha_0)=R\alpha_0\xun$,  this is, when it reaches the floor ($y=0$).

When an initial velocity is added to the particle, it slides on the frictionless
catenary until, according to Eq.~\eqref{eq:intersec-catenary}, we have
$\cosh\alpha=2-v_0^2/gR$. We conclude that the point of departure, in terms
of $\alpha$, is given by $\alpha=\ln[\varepsilon+\sqrt{\varepsilon^2-1}]$,
where $\varepsilon=2-v_0^2/gR$. From this expression, we recover the result
for zero initial velocity, $\alpha_0=\ln(2+\sqrt3)=\cos^{-1}(2)$
(or $\phi=60^\text{o}$).

\begin{figure}[!h]
\includegraphics[width=7cm]{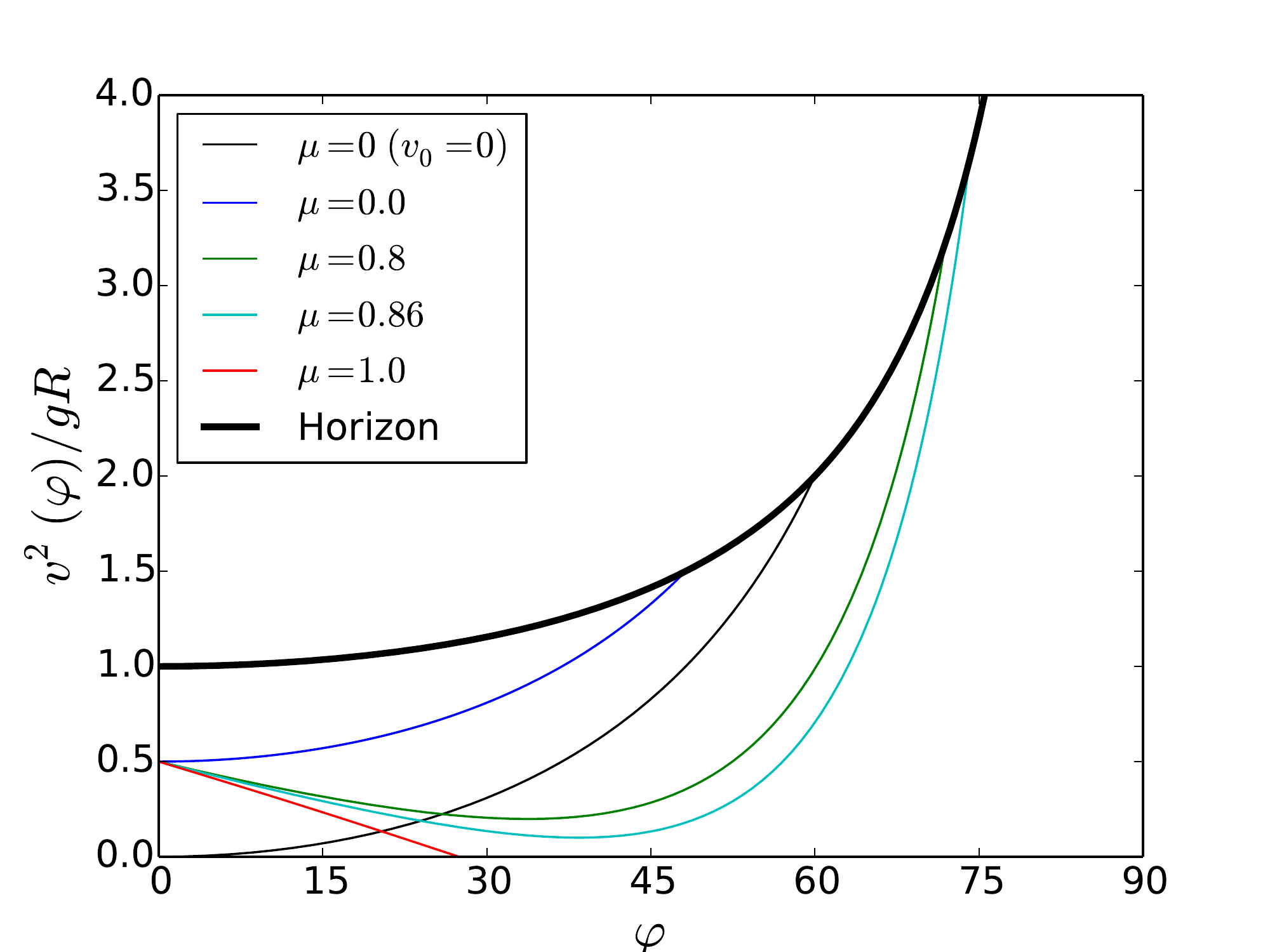}
\caption{Squared velocity as a function of $\phi$ and the horizon of the catenary.
$H(\phi)=\sec\phi$. Initial velocity $v_0^2/gR=0.5$.
For some conditions, the particle sticks to the surface, but when it does not
stop, it never touches the horizon (no departure from the surface).} 
\label{fig:horizon-catenary}
\end{figure}

In Fig.~\ref{fig:horizon-catenary} we show the evolution of $v^2(\phi)/gR$ as it
approaches the horizon of the catenary, $H(\phi)=\sec\phi$.

\subsection{Departure from a Cycloid}




%

\begin{figure}[!h]\centering
\includegraphics[width=6cm]{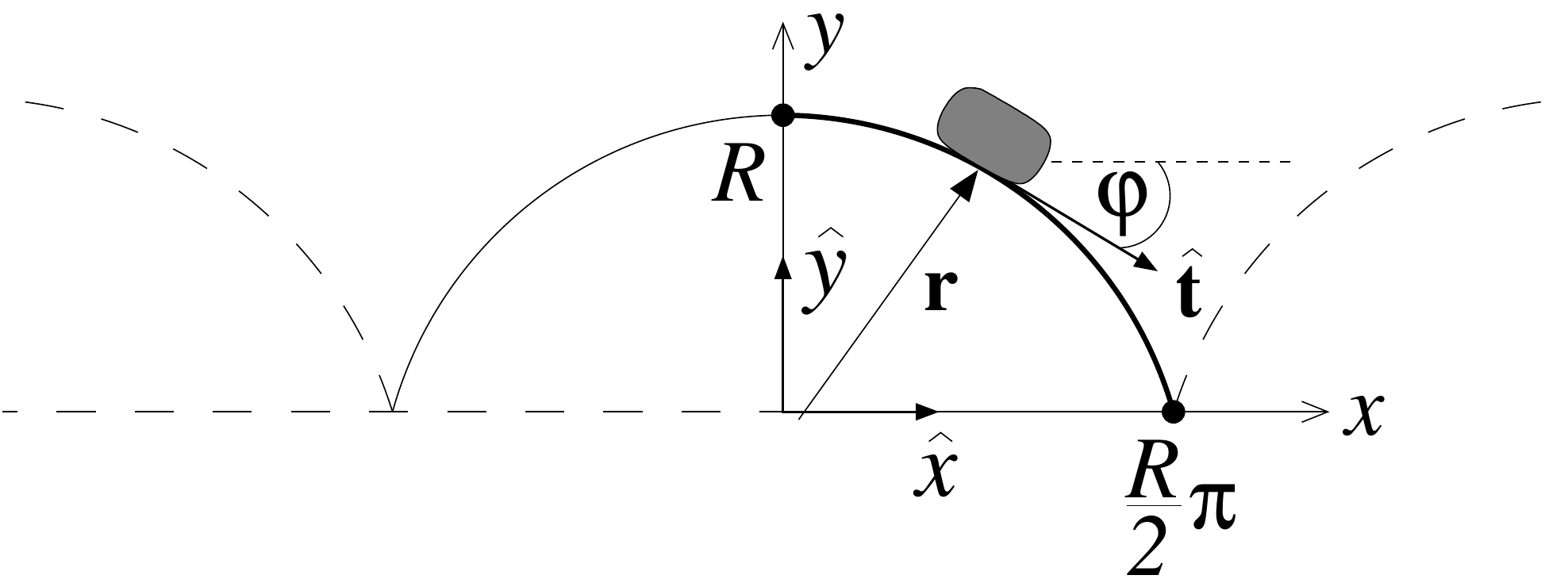}
\caption{A cycloid represented by $\v r(\alpha)=\frac R2(\alpha+\sin\alpha,1+\cos\alpha)$.}
\label{fig:cicloide}
\end{figure}

\begin{figure}[!h]
\includegraphics[width=7cm]{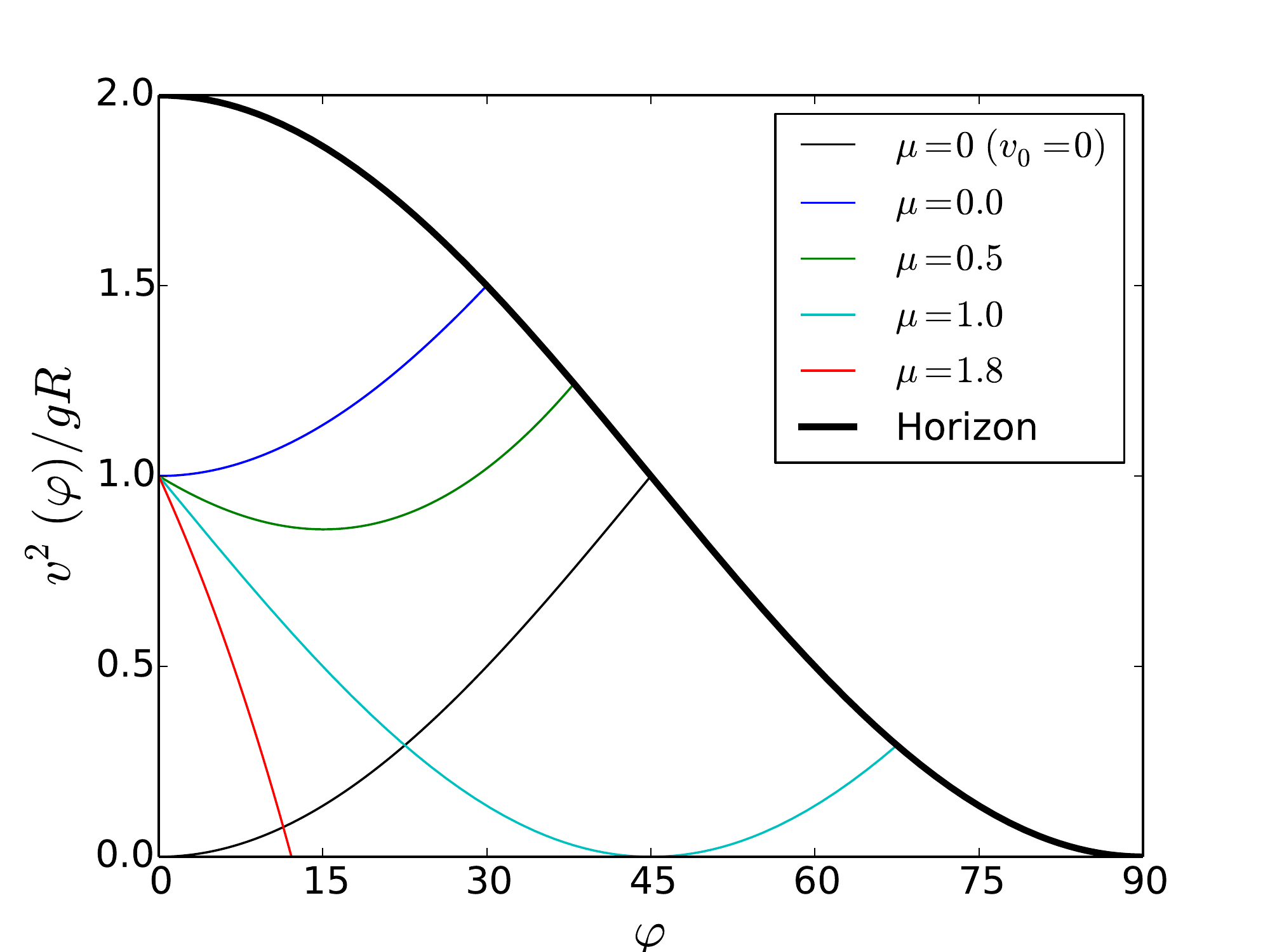}
\caption{Squared velocity as a function of $\phi$ and the horizon of the cycloid.
$H(\phi)=2\cos^2\phi$. Initial velocity $v_0^2/gR=1$.
We observe the same behaviour as in the other cases: it is always possible
to stop the particle in the surface for a big enough $\mu$.} 
\label{fig:horizon-cycloid}
\end{figure}

%
The cycloid of height $R$ can be parametrized by
$\v r(\alpha)=\frac R2(\alpha+\sin\alpha,1+\cos\alpha)$, which produces the
curve shown in Fig.~\ref{fig:cicloide}.
With this parametrization, the tangent vector is
$$
\tun=\cos(\alpha/2)\xun-\sin(\alpha/2)\yun
$$
which is equal to $\tun=\cos\phi\xun-\sin\phi\yun$. Therefore
\begin{equation}
\phi=\frac12\alpha.
\end{equation}
From here,
\begin{equation}
\kappa(\phi)=\frac1{2R}\sec\phi,
\end{equation}
which gives us the expression for the velocity:
\begin{equation}
\frac{v^2(\phi)}{gR}=
\frac{v_0^2}{gR}\e^{2\mu\phi}+4\e^{2\mu\phi}
\int_0^\phi[\sin\phi'-\mu\cos\phi']\frac{\e^{-2\mu\phi'}}{\sec\phi'}d\phi'.
\end{equation}
As the case of the circle and the parabola, this integral can be solved
analitically, since
$${\begin{split}\frac{d}{d\phi}[\e^{-2\mu\phi}(\sin\phi-\mu\cos\phi)^2]
&=\frac{\e^{-2 \mu \phi}}{ \sec\phi} (\sin\phi-\mu \cos\phi)\\
&\phantom{=} \times 2\left(1+\mu^2\right) 
\end{split}}
$$
from where we obtain
\begin{equation}
\frac{v^2(\phi)}{gR}=
\frac{v_0^2}{gR}\e^{2\mu\phi}+\frac{2(\sin\phi-\mu\cos\phi)^2-2\mu^2\e^{2\mu\phi}}{1+\mu^2}.
\end{equation}
Notice that when $\mu=1$ and $v_0^2/gR=1$, the particle stops exactly at
$\phi=\pi/4$ (see Fig.~\ref{fig:horizon-cycloid}). In the circle,  as shown by
Mungan~\cite{Mu03}, the particle reaches $\phi=\pi/4$ with zero velocity when
$\mu=1$ and $v_0^2/gR=\frac25\left(1+\sqrt2e^{-\pi/2}\right)
\approx0.52$.

The horizon of the catenary, according to Eq.~\eqref{eq:horizon-phi}, is
\begin{equation}
H(\phi)=2\cos^2\phi,
\end{equation}
which sets a maximum initial velocity of
\begin{equation}
\left(\frac{v_0^2}{gR}\right)^{(\text{max})}=H(0)=2,
\end{equation}
twice the velocity that can be given in a circle (since the curvature at
  $\phi=0$ is half that of the circle).
For $\mu=0$, the velocity is
\begin{equation}
\frac{v^2(\phi)}{gR}=
\frac{v_0^2}{gR}+2\sin^2\phi.
\end{equation}
In this case, the point of departure is satisfied when
$$
\frac{v_0^2}{gR}+2\sin^2\phi=2\cos^2\phi,
$$
therefore
$$
\cos(2\phi)=\frac{v_0^2}{2gR}.
$$
Notice that this means that $v_0=0$ implies $\phi_0=\pi/4$ ($\alpha_0=\pi/2$).
Since a cycloid is drawn from a point in the surface of a circle, which in
this case has a radius $R/2$, this situation concides with the moment in which
this circle has moved $1/4$ of the whole turn:
$$\v r(\alpha_0)=\left[\left(\frac R2\right)
 +\frac12\left(\frac R2\right)\pi\right]\xun +\left[\frac R2\right]\yun.$$
The slope in this point ($\phi=45^{\text{o}}$) is slightly smaller than in the
case of the frictionless circle, where $\phi\approx 48^{\text{o}}$.

In Fig.~\ref{fig:horizon-cycloid} we show the evolution of $v^2(\phi)/gR$ as it
approaches the horizon of the cycloid, $H(\phi)=2\cos^2\phi$.

\subsection{Phase diagram $\boldsymbol\mu$--$\boldsymbol{v_0}$}\label{sec:phase-separation}
\begin{figure}[h]
\includegraphics[width=7cm]{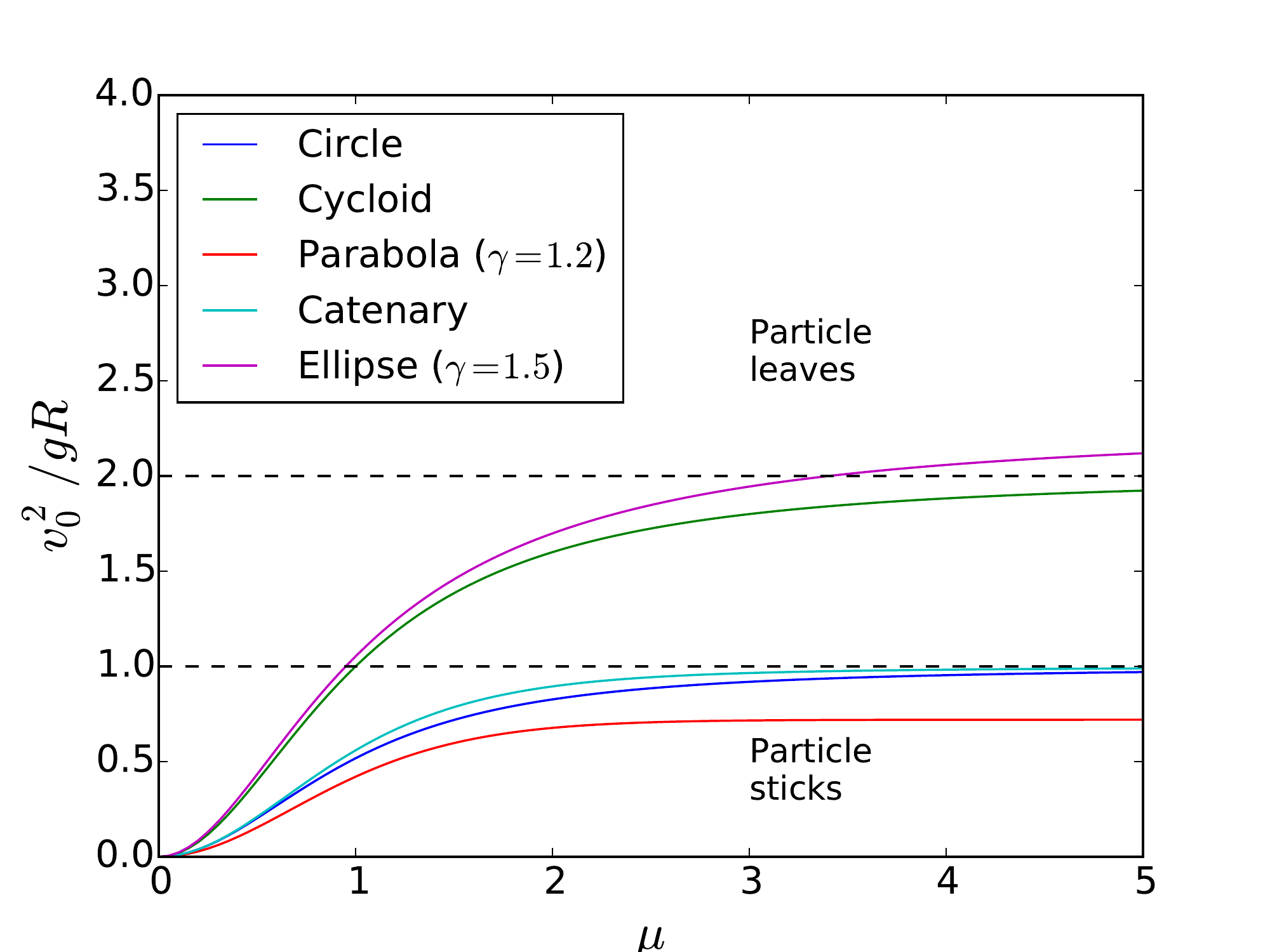}
\caption{Phase diagram separating the conditions $\mu$--$v_0$ that lead to a
particle stuck in the surface, to those conditions that allow it to move without
 stop. The curves are plots of equations~\eqref{eq:phase-circle} (circle),
\eqref{eq:phase-parabola} (parabola),~\eqref{eq:phase-cycloid} (cycloid),
\eqref{eq:phase-catenary} (catenary) and~\eqref{eq:phase-ellipse} (ellipse). As
$\mu$ increases, the velocity converges to 1 in the case of the circle and catenary,
to 2 for the cycloid. For the chosen values of $\gamma$, the velocity converges
to $\gamma^2/2=0.72$ in the case of the parabola, and to $\gamma^2=2.25$
for the ellipse.}
\label{fig:phase-diagram}
\end{figure}

As we increase $\mu$ for a particular value of $v_0>0$, there is a moment in
which the velocity splits into two branches: the first one physical (red line
in all previous figures), where the particle stops in the surface for
some critical angle $\phi_c$, and the second non-physical (not drawn in the
figures), which appears as a mathematical continuation of the function
$v^2(\phi)$ after it becomes negative. The critical angle $\phi_c$ is
obtained, as de Lange \emph{et. al}~\cite{LPM08} did for the circle, when the velocity
satisfies the following two conditions:
\begin{equation}\label{eq:conditions-mu-phi}
v^2(\phi_c)=0, \quad \left[\frac{d}{d\phi}\left(v^2(\phi)\right)\right]_{\phi_c}=0.
\end{equation}
By applying these conditions to the velocity equation~\eqref{e:ecdifv2},
we see that the left-hand side of the equation vanishes, and a general result is obtained:
\begin{equation}\label{eq:mu=tan(phi)}
\mu=-(\bun\cdot\zun)\frac{\v g\cdot\tun(s_c)}{\v g\cdot\nun(s_c)}=\tan\phi_c
\end{equation}
regardless of the sign of $(\bun\cdot\zun)=\pm1$. Here, $s_c=s(\phi_c)$.
This result is particulary meaningful, since when a particle moves in an
inclined rough plane (where $\kappa(\phi)=0$) with a friction coefficient
equal to the slope of the plane ($\mu=\tan\phi$), the particle travels at
constant velocity. If we substitute
this result ($\phi_c=\tan^{-1}\mu$) in Eq.~\eqref{eq:velocityphi-inc}, using the
conditions in Eq.~\eqref{eq:conditions-mu-phi},
we obtain $v_0$ as a function of $\mu$:
\begin{equation}\label{eq:phase-general}
  \frac{v_0^2(\mu)}{gR} =2\int_{0}^{\tan^{-1}\mu} (\mu \cos\phi'-\sin\phi') 
  \frac{\e^{-2 \mu \phi'}}{R\kappa(\phi')}\,d\phi'.
\end{equation}
This curve defines a phase diagram for the initial conditions $\mu,v_0$, that separates those
that lead to a particle that sticks in the surface from those that allow it to
move on and reach the horizon. For the circle, parabola and the cycloid this
relationship is:
\begin{eqnarray}\label{eq:phase-circle}
\frac{v_0^2(\mu)}{gR}
&=&\frac{ 2(2\mu^2-1+\e^{-2\mu\tan^{-1}\mu}\sqrt{1+\mu^2}) }{1+4\mu^2},\\\label{eq:phase-parabola}
\frac{v_0^2(\mu)}{gR}
&=&\frac{\gamma ^2}2 \left(1-(1+\mu^2)\e^{-2 \mu\tan^{-1}\mu}\right),\\\label{eq:phase-cycloid}
\frac{v_0^2(\mu)}{gR}&=&\frac{2\mu^2}{1+\mu^2},
\end{eqnarray}
respectively. As $\mu$ increases, these velocities tend to 1,
$\gamma^2/2$ and 2, respectively.
Since the integral term of the velocity cannot be
calculated explicitly neither for the catenary, nor the ellipse, the
substitution $\phi_c=\tan\mu$ also leads to non integrable expressions:
\begin{equation}\label{eq:phase-catenary}
\frac{v_0^2(\mu)}{gR}=2\int_0^{\tan^{-1}\mu}[\mu\cos\phi'-\sin\phi']
\frac{\e^{-2\mu\phi'}}{\cos^2\phi'}d\phi'
\end{equation}
for the catenary, and
\begin{equation}\label{eq:phase-ellipse}
\frac{v_0^2(\mu)}{gR}=2\int_0^{\tan^{-1}\mu}
\frac{\gamma^2[\mu\cos\phi'-\sin\phi']}{(\gamma^2\sin^2\phi+\cos^2\phi)^\frac32}\e^{-2\mu\phi'}d\phi'
\end{equation}
for the ellipse, which tends to 1 and $\gamma^2$, respectively.
As we see, these limits correspond to the
maximum initial velocity in Eq.~\eqref{eq:maxinitvel}:
\begin{equation}
\lim_{\mu\to\infty} v_0^{(\text{crit})}(\mu)=v_0^{\text{max}}.
\end{equation}
All these curves are shown in Fig.~\ref{fig:phase-diagram}.

\subsection{Curves with no point of departure}

A general behavior observed in the analysis performed is that, if the friction
coefficient does not stop the particle, the velocity always starts increasing
in some point. It is clear that a flat or decreasing horizon (constant or
increasing curvature, such as the circle, the ellipse and the cycloid) will
always be intersected by the velocity, that is, the particle will always leave
the surfaces characterized by a decreasing horizon.
But if the horizon increases, it is not clear whether the
velocity will reach the horizon or not. Some horizons do not increase fast
enough, and the velocity intersects them (such as the catenary); but some
horizons increase faster than the velocity and the intersection with it
never takes place, as we saw in the case of the parabola.

Since the behavior of the horizon
depends on the curvature, it is necessary to study different curvatures. For
a given curve of known curvature, the relationship
\begin{equation}
\frac{d\v r}{d\phi}=\frac{d\v r}{ds}\frac{ds}{d\phi}=\frac{\tun}{d\phi/ds}=\frac{\tun(\phi)}{\kappa(\phi)},
\end{equation}
where $\tun=\cos\phi\xun-\sin\phi\yun$, allows to build a parametrization
$\v r(\phi)=(x(\phi),y(\phi))$ for the curve, since it implies that
$$(x'(\phi),y'(\phi))=\left(\frac{\cos\phi}{\kappa(\phi)},-\frac{\sin\phi}{\kappa(\phi)}\right),$$
from where
\begin{subequations}
\begin{align}
x(\phi)&=x_0+\int_0^\phi \frac{\cos\phi'}{\kappa(\phi')}\,d\phi'\\
y(\phi)&=y_0-\int_0^\phi \frac{\sin\phi'}{\kappa(\phi')}\,d\phi',
\end{align}
\end{subequations}
where $x_0=0$ and $y_0=R$ represent the departure point we have used in all cases.
Notice that all the curvatures of the curves we have presented can be
expressed by $\kappa(\phi)=\cos^n\phi/R$. For example, the case of the parabola
corresponds to $n=3$ (see Eq.~\eqref{eq:curvature-parabola}),
and from the above equations we obtain $x(\phi)=R\tan\phi$,
$y(\phi)=R-R(1-\sec^2\phi/2)$, which indeed
represents the parabola $y(x)=\frac12(R-x^2/R)$.  In similar fashion,
it represents a catenary if $n=2$, the function $y(x)=R+R\ln\cos(x/R)$ if
$n=1$, a circle if $n=0$, and a cycloid if $n=-1$. The velocity intersects
the horizon $H(\phi)=\cos\phi/R\kappa(\phi)$ in these four cases, and also for
every $n<3$, since the horizon will be decreasing.

Any curve such that $H(\pi/2)=const.\neq0$ will also be intersected by the
increasing velocity in some point, meaning that the particle will departure from the
surface. This happens when $\kappa(\pi/2)=0$ and $\kappa'(\pi/2)\neq0$,
for example $\kappa(\phi)=(1-2\phi/\pi)/R$,
in which case $\lim_{\phi\to\frac\pi2}H(\phi)=-1/\kappa'(\frac\pi2)=\frac\pi2$
exists, and the horizon is a slowly increasing function. But not all curvatures
that are zero at $\phi=\frac\pi2$ produce a constant value of $H$ at this angle.
This is the case of the catenary and the parabola, whose curvature is zero at
$\phi=\frac\pi2$ and the horizon diverges at this point. However, a divergent
horizon (which is obtained when $\kappa(\frac\pi2)=\kappa'(\frac\pi2)=0$) is
still not sufficient requirement to avoid the velocity to reach it, as the case
of the catenary show us. Therefore, we observe that a sufficient condition
to avoid an intersection between the velocity and the horizon (to keep the
particle always in the curve) is to require $\kappa$ to be a decreasing function
($\kappa'(\phi)<0$) such that
\begin{equation}\label{eq:no-departure-condition}
\kappa\left(\frac\pi2\right)=\kappa'\left(\frac\pi2\right)=\kappa''\left(\frac\pi2\right)=0.
\end{equation}
We believe that it is necessary to require the monotonically decreasing condition,
since curvatures such as $\kappa(\phi)=\phi(1-2\phi/\pi)^3$, which increases
and then decreases, satisfy Eq.~\eqref{eq:no-departure-condition} and still allow
an intersection between the horizon and a the velocity.
In this way, the family of curves $\kappa(\phi)=\cos^n\phi/R$, with
$n\geqslant3$ have non-intersectable horizons, with the parabola being only
a particular case. Other examples are easy to find, such as the family of functions
$\kappa(\phi)=(1-2\phi/\pi)^n/R$ for $n\geqslant3$, in which the particle also
stays always in the surface.

We would also like to remark that the family of
functions $f(x)=R(1-\alpha x^n)$ have a curvature
$\kappa(\phi)=(\alpha n)^{\frac1{n-1}}|n-1|\cos^3(\phi)\tan^{\frac{n-2}{n-1}}(\phi)$,
which is zero at $\phi=0$ and $\phi=\frac\pi2$ for $n>2$. This means that the
horizon diverges in these two points, and despite of the fact that this
curvature satisfies the conditions in Eq.~\eqref{eq:no-departure-condition},
an intersection between the horizon and the velocity does exist. When
$1<n<2$, the curvature diverges at $\phi=0$ and $\lim_{\phi\to0}H(\phi)=0$.
This means that, according to Eq.~\eqref{eq:maxinitvel}, the maximum velocity that
can be given to the particle is zero, which in turns means that at the
slightest impulse provided to the particle, it will leave the surface. This
result has been previously explained~\cite{Ag12}, but here we have presented
it from the point of view of a quantifiable maximum initial velocity.
The cases $n=1$ and $n=2$ are the
inclined plane and the parabola, in which we already know that the particle
will remain always in the surface. When $0<n<1$, the concavity changes and
$\bun\cdot\zun=1$, which means that the particle always stays on the surface.






%
%
\section{Suggested Projects}
We have developed a general treatment to find the velocity of a particle sliding 
in an arbitrary surface with friction, providing a framework to find the point
of departure. For each surface, a maximum initial velocity is found, above
which the particle leaves the surface immediately.
We introduced the concept of the {\it horizon} of a curve, which has proven
to be fundamental to find the point of departure, given by the intersection
between the velocity and the horizon.
A summary of all the curves analyzed is shown in Table~\ref{tab:summary}.
\begin{table*}\centering
\begin{tabular}{| >{\centering\arraybackslash}m{2cm} | >{\centering\arraybackslash}m{3cm} | >{\centering\arraybackslash}m{3cm}|>{\centering\arraybackslash}m{3cm}|>{\centering\arraybackslash}m{3cm}|>{\centering\arraybackslash}m{3cm}|}\hline
& Circle & Ellipse & Parabola & Catenary & Cycloid \\\hline
$\v r(\alpha)$
& $(R\sin\alpha,R\cos\alpha)$
& $(R\gamma\sin\alpha,R\cos\alpha)$
& $(R\gamma\alpha,R(1-\alpha^2))$
& $(R\alpha,R(2-\cosh\alpha))$
& $\frac R2(\alpha+\sin\alpha,1+\cos\alpha)$
\\\hline
$\phi(\alpha)$
& $\phi=\alpha$ 
& $\tan\phi=(1/\gamma)\tan\alpha$ 
& $\tan\phi=2\alpha/\gamma$
& $\tan\phi=\sinh\alpha$
& $\phi=\frac12\alpha$
\\\hline
$\kappa(\phi)$
& $1/R$
& $\frac{(\gamma^2\sin^2\phi+\cos^2\phi)^\frac32}{\gamma^2R}$
& $\frac{2}{\gamma^2R}\cos^3\phi$
& $\frac1R\cos^2\phi$
& $\frac1{2R}\sec\phi$
\\\hline
$\dfrac{v^2(\phi)}{gR}$
& $\frac{v_0^2}{gR}\e^{2\mu\phi}+\frac{(2-4\mu^2)\left(e^{2\mu\phi}-\cos\phi\right)}{1+4\mu^2}$ $-\frac{6\mu\sin\phi}{1+4\mu^2}$
& $\frac{v_0^2}{gR}\e^{2\mu\phi}+\frac2R\e^{2\mu\phi}\int_0^\phi[\sin\phi'-\mu\cos\phi']\frac{\e^{-2\mu\phi'}}{\kappa(\phi')}d\phi'$
& $\frac{\gamma^2}{2}\sec^2\phi-\left(\frac{\gamma^2}{2}-\frac{v_0^2}{gR}\right)\e^{2\mu\phi}$
& $\frac{v_0^2}{gR}\e^{2\mu\phi}+2\e^{2\mu\phi}\int_0^\phi[\sin\phi'-\mu\cos\phi']\frac{\e^{-2\mu\phi'}}{\cos^2\phi'}d\phi'$
& $\frac{v_0^2}{gR}\e^{2\mu\phi}+\frac{2(\sin\phi-\mu\cos\phi)^2}{1+\mu^2}-\frac{2\mu^2\e^{2\mu\phi}}{1+\mu^2}$
\\\hline
$\dfrac{v^2(\phi)}{gR}$ for $\mu=0$
& $\frac{v_0^2}{gR}+2(1-\cos\phi)$
& $\frac{v_0^2}{gR}+2-\frac{2\cos\phi}{\sqrt{\gamma^2\sin^2\phi+\cos^2\phi}}$
& $\frac{v_0^2}{gR}+\frac{\gamma^2}{2}\tan^2\phi$
& $\frac{v_0^2}{gR}+2(\sec\phi-1)$
& $\frac{v_0^2}{gR}+2\sin^2\phi$
\\\hline
$\left(\dfrac{v_0^2}{gR}\right)^{(\text{max})}$
& 1
& $\gamma^2$
& $\frac12\gamma^2$
& 1
& 2
\\\hline
Horizon
& $\cos\phi$
& $\frac{\gamma^2\cos\phi}{(\gamma^2\sin^2\phi+\cos^2\phi)^{3/2}}$
& $\frac{\gamma^2}{2}\sec^2\phi$
& $\sec\phi$
& $2\cos^2\phi$
\\\hline
\end{tabular}
\caption{Summary of the curves. }
\label{tab:summary}
\end{table*}

Here we propose a set of possible projects with different levels of complexity.

\subsection{Power Laws}
Surfaces described by a decreasing function 
$f\colon\mathbb R\to\mathbb R$, have a natural parametrization
$\v r(x)=(x,f(x))$, such as $f(x)=-\alpha x^k$. Find their
curvature and demonstrate that for $k>2$ it is not possible
to throw the particle from the origin with any bounded initial
velocity such that it immediately leaves the surface.
Demonstrate that any curve with an inflection point at the
origin also shows this behavior.


\subsection{Nephroid}
\begin{figure}[!h]
\includegraphics[width=5cm]{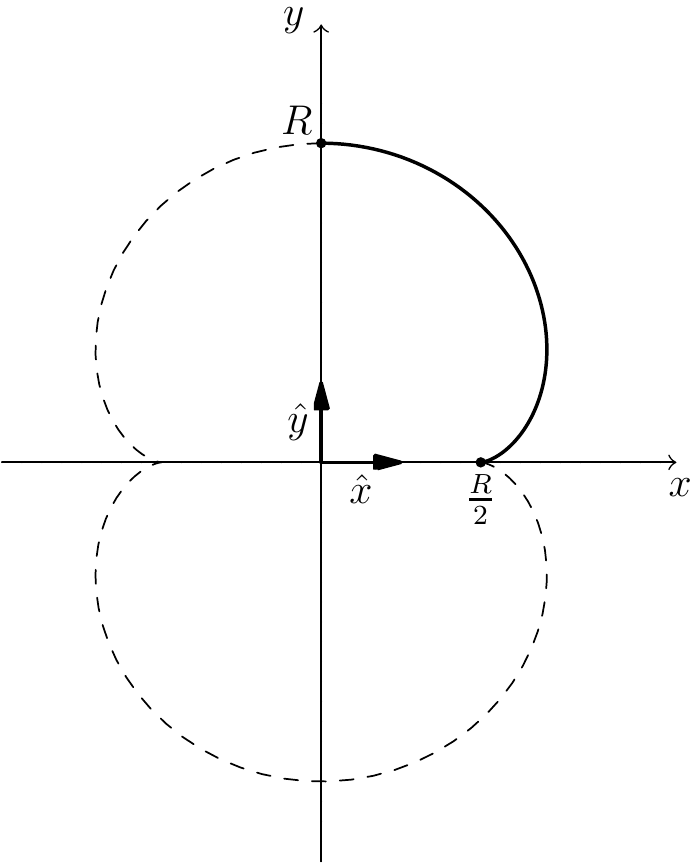}
\caption{A nephroid.} 
\label{fig:nephroid}
\end{figure}

Another interesting, famous curve is the nephroid. A possible parametrization
for this curve is
\begin{equation}
\v r(\alpha)= \left(\frac R4(3\sin(\alpha)+\sin(3\alpha)), \frac R4(3\cos(\alpha)+\cos(3\alpha)) \right),
\end{equation}
which we show in Fig.~\ref{fig:nephroid}.
Obtain the velocity of the
particle in terms of the tangential angle $\varphi$, the horizon of the curve and the
maximum initial velocity. Plot the velocity as a function of $\varphi$ for different
initial velocities and friction coefficients. What can be said about the departure angle?

\subsection{Other forces}
A dragging force $\v F_D=-\beta v^2(s)\tun$, where $\beta$ is a constant and
$v$ the velocity of the particle, can be incorporated to the model.
This force competes with the friction. Show that the equations of motion
in this case are
%
\begin{align}
  m v(s) \frac{d v(s)}{d s} &= m\ngr{g} \cdot \tun(s) - \mu N(s)
 - m \beta v^{2}(s) \\
  m \kappa(s) v^{2}(s) &= m\v g \cdot \nun(s) + (\bun\cdot\zun)N(s) .
\end{align}
%
Eliminate the normal force from the equations, and obtain
\begin{equation}
  \frac{\mathrm{d}}{\mathrm{d} s} \bigl( v^{2}(s) \bigr) +
  2 \left(\beta\pm\mu \kappa(s)\right)
 v^{2}(s) = 2 (\v g\cdot \tun(s) \pm \mu \v g\cdot \nun(s)).
\end{equation}
%
Note that this equation already appears in the classical
book of Appel~\cite{Ap41}, Chap. XII, \S~251. Show that
the the integrating factor that solves this equation is
$\exp\left[2\beta s(\phi)\pm2\mu(\phi-\phi_0)\right]$.
Following the treatment presented in this paper, analyze
the case of a particle sliding on a circle with friction and a dragging
force and compare to the case presented in~\ref{sec:circle}.

\section{Conclusions}
We have found analytical expressions for the velocity of the particle when
it slides on a circle, a parabola and on a cycloid, including the effects
of friction, explaining in detail
why a particle in a parabola does never leave that surface.
Explicit departure angles have been found for frictionless surfaces
when the particle starts from rest: $\phi=\arccos(2/3)\approx 48^{\text o}$
for the circle, $\phi=45^{\text o}$ for the cycloid, and $\phi=60^{\text o}$
for the catenary, an angle that coincides with the moment the particle reaches
the floor. In the roughless ellipse, we do not have a determined departure angle, since
it depends on the parameter $\gamma$. The departure angle decreases as $\gamma$
increases, and it corresponds to the circle when $\gamma=1$.
For concave upwards surfaces, there is no departure angle.

Two important, general results obtained are the relationship between the
critical angle $\varphi_c$ and the friction coefficient $\mu$, given by
Eq.~\eqref{eq:mu=tan(phi)}, and, as a consequence, the Eq.~\eqref{eq:phase-general}.
This equation relates the friction coefficient with the initial velocity and
allows us to build the phase diagram in
Fig.~\ref{fig:phase-diagram}, which separates the initial
conditions $\mu$--$v_0$ that lead to a particle stuck on the surface to those
conditions that allow it to move until it takes off (if the intersection with
the horizon exists).

Regarding the maximum velocity that can be given initially to the particle,
we can say that if the particle is pushed from the top of a curve with initially
horizontal tangent
($\nun(0)=-\yun$), according to Eq.~\eqref{eq:maxinitvel} we have
$$\left(\frac{v_0^2}{g}\right)^{(\text{max})}
=\frac{\v g\cdot\nun(0)}{g\kappa(0)}=\frac{1}{\kappa(0)},$$
which means that any curve that is locally flat at the top ($\kappa(0)=0$)
has infinite maximum initial velocity.
In other words, if the particle is kicked from a point of zero curvature, there
is no initial velocity for which the particle leaves the surface immediately.

Finally, although a formal demonstration is still lacking,
we have found a general criteria ($\kappa'(\phi)<0$
plus Eq.~\eqref{eq:no-departure-condition}) to discriminate in which curves the
particle will always stay in the surface, and in which curves the particle
takes off in some point. We have shown families of functions in which the
particle never takes off
($\kappa(\phi)=(1-2\phi/\pi)^n/R$, $\kappa(\phi)=\cos^n\phi/R$),
and we have given the tools to analyze the particular
case of power laws, in which the particle exhibits all the behaviors,
depending on the exponent:
it leaves the surface immediately, in some point on the surface, or never
leaves the surface. In addition, we have concluded that if the particle is
initially kicked from a point of zero curvature (like an inflection point),
there is no initial velocity for which the particle leaves the surface
immediately.

\begin{acknowledgments}
F.G.-C. acknowledges CONICYT PhD fellowship No. 201090712 short-term fellowship
from Universidad de Chile. G.G. thanks project Anillo ACT-1115.
\end{acknowledgments}


\newpage

\bibliography{friction}
\bibliographystyle{naturemag}

\end{document}